\documentclass[12pt]{iopart}
\usepackage{graphicx}
\usepackage{bm}
\usepackage{wick}

\newcommand{\lan }{\langle}
\newcommand{\ran }{\rangle}
\newcommand{\beq }{\begin{eqnarray}}
\newcommand{\eeq }{\end{eqnarray}}

\newcommand{\rsolid}{
\put(0,3.1){$\!\!\!$ \vector(1,0){18}
} \quad}
\newcommand{\lsolid}{
\put(0,3.1){ ~~~\vector(-1,0){18} } \quad}
\newcommand{\rdash}{
\put(0,3.1) {$\!\!$\line(1,0){3}  }
\put(8,3.1) {$\!\!$\line(1,0){3}  }
\put(15,3.1){$\!\!$\vector(1,0){5} } \quad}
\newcommand{\ldash}{
\put(0,3.1){\,\vector(-1,0){5} }
\put(8,3.1) {\,$\!\!$\line(1,0){3}  }
\put(15,3.1) {\,$\!\!$\line(1,0){3}  }
 \quad}

\usepackage{iopams}
\begin{document}

\title[
{Semiclassical Theory for Universality in Quantum Chaos with
Symmetry Crossover}
]
{Semiclassical Theory for Universality in Quantum Chaos with
Symmetry Crossover}

\author{Keiji Saito$^{1,2,*}$, Taro Nagao$^3$, Sebastian M\"uller$^4$, 
and \protect \\ 
Petr Braun$^{5,6}$ }
\address{$^1$Graduate School of Science,
University of Tokyo, 113-0033, Japan \\
$^2$ CREST, Japan Science and Technology (JST),
Saitama, 332-0012, Japan \\
$^3$Graduate School of Mathematics,
Nagoya University, Chikusa-ku, Nagoya 464-8602, Japan \\
$^4$Department of Mathematics, University of Bristol, Bristol BS8 1TW,
United Kingdom \\
$^5$Fachbereich Physik, Universit\"at Duisburg-Essen,
47048 Duisburg, Germany \\
$^6$
Institute of Physics, Saint-Petersburg University, 198504 Saint-Petersburg,
Russia \\
}
\ead{$^*$saitoh@spin.phys.s.u-tokyo.ac.jp}

\begin{abstract}
We address the quantum-classical correspondence for chaotic systems with a 
crossover between symmetry classes.
We consider the energy level statistics of
  a classically chaotic system in a weak magnetic field. 
The generating function of spectral correlations is calculated by
  using the semiclassical periodic-orbit theory. An explicit
  calculation up to the second order, including the non-oscillatory
  and oscillatory terms, agrees with the prediction of random
  matrix theory. Formal expressions of the higher order terms are also
  presented. The nonlinear sigma (NLS) model of random matrix theory,
  in the variant of the Bosonic replica trick, is also analyzed for
  the crossover between the Gaussian orthogonal ensemble and Gaussian
  unitary ensemble. The diagrammatic expansion of the NLS model is
  interpreted in terms of the periodic orbit theory.
\end{abstract}

\pacs{05.45.Mt}
\maketitle

\section{Introduction}
The quantum-classical correspondence in a chaotic system is a
longstanding problem. In a quantum system, chaos does not exist
in the sense of the "sensitivity to the initial conditions", while in a
classical system, it exists and a characterization is given by a
positive Lyapunov exponent. A signature of chaos in a quantum
system can be seen in the energy level statistics, which shares
the same behavior with the eigenvalue statistics of random
matrices~\cite{haake}. According to a well-known conjecture by
Bohigas, Giannoni, and Schmit~\cite{bgs}, the energy level
statistics in classically chaotic systems with and without
time-reversal symmetry are universally described by the Gaussian
orthogonal ensemble (GOE) and Gaussian unitary ensemble (GUE),
respectively.

Recently a justification of this conjecture has been given for
the spectral form factor, in terms of the semiclassical periodic-orbit
theory\cite{muller}. It is based on Berry's diagonal approximation~\cite{berry}
for the first order term and the periodic-orbit pairs with an encounter~\cite{sr}
for the second order term. Higher order terms are also calculated by using the
periodic pairs with many encounters. If a very weak magnetic field is
applied to a classically chaotic system, a crossover between the GOE and GUE
universality classes is realized. The level statistics in this domain is
described by Pandey and Mehta's two matrix model~\cite{PM83}. The periodic-orbit
theory is also applicable to explain the behavior in this GOE-GUE crossover
domain~\cite{bgoas, nagaosaito2003, tr, saitonagao2006, nagaohaake2007}.
In this approach, a stochastic behavior of the periodic orbits plays a
crucial role. Similar ideas also give evaluations
of
GUE-GUE\cite{nagaohaake2007, sieber2007},
GOE-GOE\cite{sieber2007}, GSE-GSE\cite{nagaosaito2007}
and GOE-GSE\cite{nagaosaito2007} form factors.

Moreover a recent progress on this topic has made it possible to
reproduce the oscillatory terms in the expansion of the GOE and GUE
spectral correlations~\cite{heusler2007}. In this new scheme, a periodic
orbit theory is constructed for the generating function which yields
the spectral correlation. The generating function is defined as
\begin{eqnarray}
Z (\epsilon_A , \epsilon_B , \epsilon_C , \epsilon_D )&=& \left\langle
{
\det \left( E_C^+ - H  \right) ~\det \left( E_D^- - H  \right)
 \over
\det \left( E_A^+ - H  \right) ~\det \left( E_B^- - H  \right)
 }
\right\rangle.
\label{def_of_gene}
\end{eqnarray}
where $H$ is the Hamiltonian. Here $E_{A,C}^{+}$ and $E_{B,D}^{-}$ are given by
$E_{A,C}^{+}
 = E + {\epsilon_{A,C}^{+} \over 2\pi \bar{\rho}}$
and $E_{B,D}^{-} = E + {\epsilon_{B,D}^{-} \over 2\pi
\bar{\rho}}$ with 
the local average of the level density $\bar{\rho}$, where the offsets from the center energy $E$ are taken with an 
imaginary part as
$\epsilon^{\pm} = \epsilon \pm i \gamma$
($\gamma > 0$) to ensure the convergence. The generating function
(\ref{def_of_gene}) yields the spectral correlation  function
$C(\epsilon)$ as \beq \left. {\partial^2 Z \over \partial
\epsilon_A \partial \epsilon_B} \right|_{||} &=& {1\over
(2\pi\bar{\rho})^2} {\rm Tr}\left( {1\over E + {\epsilon \over 2
\pi \bar{\rho} } + i\gamma -H }\right) {\rm Tr}\left( {1\over E -
{\epsilon \over 2 \pi \bar{\rho} } - i\gamma -H }\right)
\nonumber \\
&=&{1\over 2} C(\epsilon ) + {1\over 4} , \label{part2} \eeq where
the symbol $||$ denotes the identification  $\epsilon_A^+ =
\epsilon_C^+ = \epsilon^+, \epsilon_B^- = \epsilon_D^- =
-\epsilon^+$. A crucial idea of the new scheme is counting all the
diagrams in terms of the pseudo orbits, which correspond to the
terms appearing in a field theoretical treatment of random
matrices.

In this paper, we extend this new scheme to the case with a weak
magnetic field, which describes the GOE-GUE crossover. We combine
the ideas of Refs.\cite{heusler2007} and \cite{saitonagao2006},
and compare the results of periodic-orbit and random matrix
theories. The random matrix theory prediction for the
generating function is known to be \cite{Zirnbauer},%
 \beq Z
(\epsilon_A , \epsilon_B , \epsilon_C , \epsilon_D )&=& Z^{(1)} +
Z^{(2)} , \eeq where \beq Z^{(1)} &=& e^{i (\epsilon_A -
\epsilon_B - \epsilon_C + \epsilon_D )/2} {     (\epsilon_A -
\epsilon_D )(\epsilon_C - \epsilon_B ) \over (\epsilon_A -
\epsilon_B )(\epsilon_C - \epsilon_D ) } ~[ 1 + Z^{(1)}_{{\rm
off}} ],
  \\
 Z^{(1)}_{{\rm off}} &= &
{ \epsilon_{AC}  \epsilon_{BD}  \over 4} \int_{1}^{\infty} dp ~
p^{-1} \ e^{i(p-1)\epsilon_{AB}/2 -2s p^2} \int_{0}^{1} dq ~ q \
e^{-i(q-1)\epsilon_{CD}/ 2 +2s q^2}
\nonumber \\
&=&{2\over i (\epsilon_{AB} + i8s )}- {2\over i (\epsilon_{AD} +
i8s )}- {2\over i (\epsilon_{CB} + i8s )}+
{2\over i (\epsilon_{CD} + i8s )}   \nonumber \\
&+& 16s \epsilon_{AC} \epsilon_{BD} \left[ {1\over (\epsilon_{CD}
+ i8s)^2
        (\epsilon_{AD} + i8s )
        (\epsilon_{CB} + i8s)   } \right. \nonumber \\
&& - \left. {1\over (\epsilon_{AB} + i8s)^2
        (\epsilon_{AD} + i8s)
        (\epsilon_{CB} + i8s)   }
\right] +\cdots \eeq%
 with $\epsilon_{\alpha \beta} =
\epsilon_{\alpha} - \epsilon_{\beta}$. The function $Z^{(2)}$
is obtained from the function $Z^{(1)}$ by exchanging
$\epsilon_C$ and $\epsilon_D$. The parameter $s$ controls the
symmetry: the GOE (GUE) limit corresponds to $s\to 0 ~(s\to\infty
)$. In this paper, we explicitly calculate the generating function
up to the second order using the periodic-orbit theory, and
confirm the agreement with the above prediction. Moreover, in
order to see the correspondence of the higher order terms, we
examine the nonlinear sigma (NLS) model in the crossover domain,
and interpret the diagrammatic expansion of the NLS model in terms
of periodic pairs.
\section{Semiclassical Expression of the Generating Function}

\subsection{Magnetic Action}
 We consider a bounded quantum system with $f$ degrees of freedom
whose corresponding classical
dynamics is chaotic.
We assume that the system has a gauge potential due to a magnetic field $B$.
Let us denote the energy by $E$ and each phase
space point by a $2f$ dimensional vector ${\bm x} = ({\bm q},{\bm p})$,
where $f$ dimensional vectors ${\bm q}$ and ${\bm p}$ specify the position
and momentum, respectively. In the semiclassical limit $\hbar \to
0$, the trace formula for the energy-level density implies
\begin{eqnarray}
{\rm Tr}\left( E^{+} - H \right)^{-1}
 & \sim  &-i\pi \bar{\rho} - {i\over \hbar } \sum_{a} T_a F_a
e^{{i\over \hbar} [ S_a (E^{+}) + \Theta_a (B)] } , \label{trace_formula}
\end{eqnarray}
where $H$ is the Hamiltonian. The first term in the right hand side 
gives the local average 
of the level density ${\bar \rho}$
and the second term describes the fluctuation around the
average.
The local average of the level density is equal to the number
of Planck cells inside the energy shell
\begin{eqnarray}
\bar{\rho} = \frac{\Omega(E)}{(2 \pi \hbar)^f},
\end{eqnarray}
where $\Omega(E)$ is the volume of the energy shell.
The function $S_a$ is the classical action (including the
    Maslov phase) of the periodic orbit $a$ with a period
    $T_a = dS_a/dE$, $F_a$ is the stability amplitude
    and $\Theta_a$ is the magnetic action.

We consider the following two possible situations.  One is the
case that the system has an Aharonov-Bohm type gauge potential,
where the gauge potential exists although the magnetic field is
not directly applied to the system. Then the classical dynamics is
time-reversal invariant, while the quantum counterpart is not. 
Another situation is the case that the magnetic field
is directly applied to the particle, but it is sufficiently weak
such that the cyclotron radius is much larger than the system size
and thus the presence of the magnetic field does not significantly
change $\Omega(E)$. 
In this case the impact of the magnetic 
field on the orbits can be neglected. In both cases the magnetic field 
still influences the behavior of the system through its contribution to the 
phase in (6). Time-reversed classical dynamics must be taken into account 
to derive the form factor in the crossover domain; it even becomes 
nontrivial to derive the form factor in the GUE limit.

The magnetic action
$\Theta_{a}(B)$ is a function of the magnetic field and is defined
as
\begin{eqnarray}
\Theta_{a}(B) = B \int_{a} {\bm A}({\bm q}) \cdot {\rm d}{\bm q}
= B \int g_{a}(t) \ {\rm d}t, \ \ \
g_{a}(t) = {\bm A}({\bm q}_{a}) \cdot
\frac{{\rm d}{\bm q}_{a}}{{\rm d}t},
\end{eqnarray}
where ${\bm A}({\bm q})$ is the gauge potential which generates
the unit magnetic field and ${\bm q}_{a}(t)$ describes a classical
motion in the configuration space along the orbit $a$.
Since the classical dynamics is chaotic, successive changes
of the velocity can be regarded as independent events. Hence, if the time
$T$ elapsed on an orbit $a$ is sufficiently large, the statistics
of $g_{a} (t)$ can be regarded as that of the Gaussian white noise
satisfying the correlation $\lan \lan g_{a} (t) g_{a }(t') \ran
\ran = 2D \delta (t-t')$. The Gaussian average $\lan\lan
\cdots\ran\ran$ is computed as a functional integral
 \beq
\label{gaussian_weight} \lan \lan \ F[g_{a}] \ \ran \ran =
\frac{\displaystyle \int {\cal D}g_{a} \ {\rm exp}\left[
-\frac{1}{4 D} \int_0^T dt \{ g_{a}(t) \}^2 \right] F[g_{a}]}
{\displaystyle \int {\cal D}g_{a} \ {\rm exp}\left[ -\frac{1}{4 D}
\int_0^T dt \{ g_{a}(t) \}^2 \right]}. \eeq
 It is known that this
averaging yields the non-oscillatory terms of the spectral
correlation~\cite{saitonagao2006, nagaohaake2007}.

\subsection{Pseudo-Orbit Representation of the Generating Function}
Utilizing Eq.(\ref{trace_formula}), we can rewrite
$\det(E^{+} - H )$ as
\beq
\det(E^{+} - H ) &\propto & \exp\left\{ \int^{E^{+}} d E' \
{\rm Tr} \left( {1\over E' - H }\right)  \right\} \nonumber \\
& \propto & \exp \left[ -i\pi \bar{N} (E^+ ) - \sum_a F_a
e^{{i\over \hbar} \{ S_a (E^{+}) + \Theta_a (B) \} } \right].
\label{a_expression} \eeq%
Expanding the exponential we obtain an equivalent expression in
terms of the pseudo orbits~\cite{berry_keating90}:%
 \beq \det(E^{+}
- H ) &\propto & e^{-i\pi \bar{N} (E^+ ) } \left\{ 1 - \sum_a F_a
e^{{i \over \hbar} (S_a + \Theta_a  )} \right.
\nonumber \\
&& \left.
+{1\over 2!}\sum_{a,a'}F_a F_{a'}
 e^{{i \over \hbar} (S_a + \Theta_a )} e^{{i \over \hbar}
(S_{a'} + \Theta_{a'} ) } - \cdots
\right\} \nonumber \\
&=& e^{ -i\pi \bar{N} (E^+ ) } \sum_A F_A (-1)^{N_A} e^{{i\over
\hbar} \{S_A (E^{+}) + \Theta_A (B) \} } , \label{ex1} \eeq%
 where
$A$ is the index of a pseudo orbit, which consists of component
orbits, and $N_A$ is the number of the component orbits. The
functions $S_A$ and $\Theta_A$ are the sums of the mechanical
 and magnetic actions of the component orbits,
respectively, and $F_A$ is the product of the stability
amplitudes. The corresponding factor $\det(E^{-} - H )$ is given
by the complex conjugation of $\det(E^{+} - H )$ as \beq \det(
E^{-} - H ) \propto e^{ i\pi \bar{N} (E^- ) } \sum_A F_A^{\ast}
(-1)^{N_A} e^{-{i\over \hbar} \{ S_A (E^{-}) + \Theta_A (B) \} },
\label{ex2} \eeq where an asterisk means a complex conjugate.

Berry and Keating ~\cite{berry_keating90} postulated that
Eq.(\ref{ex1}) and Eq.(\ref{ex2}) become identical in the
limit $E^{+},E^{-}\to E$ and found a duality relation between the
contribution of the pseudo-orbits with a duration larger than a
half of the Heisenberg time $T_H = 2\pi\hbar\bar{\rho}$ and those
with a duration shorter than $T_H /2$. Thus the sum over the long
orbits can be treated as the complex conjugate of the sum over the
short orbits. This leads to the so called Riemann-Siegel
lookalike formula %
 \beq \det(E - H ) \propto e^{ -i\pi \bar{N} (E ) }
\sum_{T_A < T_H /2} F_A (-1)^{N_A} e^{{i\over \hbar} \{ S_A
(E) + \Theta_A (B) \} } + c.c. , \label{riemann_siegel} \eeq%
This formula has an analogy in the corresponding expression of
Riemann's zeta function.

Using Eqs.(\ref{a_expression}) and (\ref{ex1}), we write the
determinants of the inverse matrices as
\beq
\det( E^+ - H )^{-1} & \propto &
\exp \left[
i \pi \bar{N} (E^+ ) + \sum_a F_a
e^{ {i \over \hbar }
\{ S_a (E^+ ) + \Theta_a (B) \} } \right]
\nonumber \\
&=&
e^{ i\pi \bar{N} (E^+ ) }
\sum_A F_A
e^{{i\over \hbar} \{ S_A (E^{+}) + \Theta_A (B) \} }  , \label{inv1}\\
\det( E^- - H )^{-1} & \propto &
\exp \left[
-i \pi \bar{N} (E^- ) + \sum_a F_a^{\ast}
e^{ -{i\over\hbar } \{ S_a (E^- ) + \Theta_a (B) \} } \right]
\nonumber \\
&=& e^{ -i\pi \bar{N} (E^- ) } \sum_A F_A^{\ast} e^{-{i\over
\hbar} \{ S_A (E^{-}) + \Theta_A (B) \} }  . \label{inv2} \eeq
Inserting these expressions directly into the definition of the
generating function, we obtain a semiclassical expression :%
 \beq
Z&=& \left\langle
 \left[ e^{-i\pi \bar{N}(E_C )} \sum_{T_C < T_H / 2} F_C (-1)^{N_C}
e^{{i\over\hbar} \{ S_C (E_C ) + \Theta_C \} } + c.c. \right]
\right. \nonumber
\\ &\times&
\left[ e^{i\pi \bar{N}(E_D)} \sum_{T_D < T_H / 2} F_D^{\ast}
(-1)^{N_D}
e^{-{i\over\hbar} \{ S_D (E_D ) + \Theta_D \} } + c.c. \right] \nonumber \\
&\times&
e^{i\pi \bar{N}(E_A^+ )} \sum_{A} F_A
e^{{i\over\hbar} \{ S_A (E_A^+  ) + \Theta_A \} }  \nonumber \\
&\times & \left.
e^{-i\pi \bar{N}(E_B^- )} \sum_{B} F_B^{\ast}
e^{-{i\over\hbar} \{ S_B (E_B^- ) + \Theta_B \} }
\right\rangle.
\label{zexp1}
\eeq%
We used the Riemann-Siegel lookalike formula
(\ref{riemann_siegel}) and its complex conjugate for the numerator
and the expressions (\ref{inv1})-(\ref{inv2}) for the denominator.
Because the expression (\ref{riemann_siegel}) has two parts,
Eq.(\ref{zexp1}) in total has four parts, each of which has the
factor $e^{\pm i \pi [ {\bar N} \{ E + \epsilon_A /
(2\pi\bar{\rho} ) \} - {\bar N} \{ E - \epsilon_B /
(2\pi\bar{\rho} ) \} ] } $ or $e^{\pm i \pi [ {\bar N} \{ E +
\epsilon_A / (2\pi\bar{\rho} ) \} + {\bar N} \{ E - \epsilon_B /
(2\pi\bar{\rho} ) \} ] }$. However each part with the latter
factor has no contribution because it is expected to vanish after
averaging over $E$ due to a rapid oscillation.

Eventually we obtain two  parts for $Z$~\cite{keating_muller} as
\beq Z &=& Z^{(1)} + Z^{(2)}, \eeq where the first term $Z^{(1)}$
is written as%
 \beq Z^{(1)} &=& \sum_{A,B,C,D \atop (T_C,T_D <
T_H/2 )}
e^{i(\epsilon_A^+ - \epsilon_B^- - \epsilon_C + \epsilon_D)/2 } \nonumber \\
&& \times
\Bigl\langle
F_A F_B^{\ast} F_C F_D^{\ast} (-1)^{N_C + N_D}
e^{{i\over \hbar} \left( S_A (E) - S_B (E) + S_C (E) - S_D (E)\right) }
\Bigr\rangle
\nonumber \\
&&\times \langle \langle e^{{i\over \hbar } (\Theta_A - \Theta_B +
\Theta_C - \Theta_D )}\rangle \rangle ~ e^{{i\over T_H} (T_A
\epsilon_A^+ - T_B \epsilon_B^- + T_C \epsilon_C - T_D
\epsilon_D)}. \label{pseudo_orbits_z1} \eeq%
 Here the symbol
$\langle\langle ... \rangle\rangle$ implies the average over the
weight Eq.(\ref{gaussian_weight}). To derive
Eq.(\ref{pseudo_orbits_z1}), we used the expansion %
 $\bar{N}(E \pm
\epsilon/2\pi\bar{\rho}) \sim \bar{N}(E) \pm \epsilon/2\pi$,%
 and %
$S(E \pm \epsilon/2\pi\bar{\rho}) /\hbar \sim S(E)/\hbar \pm T
\epsilon / T_H$. The term $Z^{(2)}$ is obtained from $Z^{(1)}$
by exchanging the variables $\epsilon_C$ and $\epsilon_D$ as
 \beq
Z^{(2)} (\epsilon_A , \epsilon_B , \epsilon_C , \epsilon_D ) &=&
Z^{(1)} (\epsilon_A , \epsilon_B ,  \epsilon_D , \epsilon_C ) .
\eeq
 Utilizing these expressions, we write the second derivatives
of $Z$, which gives the spectral correlation $C(\epsilon)$, as%
\beq
\left. {\partial^2 Z \over \partial \epsilon_A \partial
\epsilon_B} \right|_{||} &=& \left. {\partial^2 Z^{(1)} \over
\partial \epsilon_A \partial \epsilon_B} \right|_{||} + \left.
{\partial^2 Z^{(2)} \over \partial \epsilon_A \partial \epsilon_B}
\right|_{||} = \left. {\partial^2 Z^{(1)} \over \partial
\epsilon_A \partial \epsilon_B} \right|_{||} + \left. {\partial^2
Z^{(1)} \over \partial \epsilon_A \partial \epsilon_B}
\right|_{\times} , \eeq%
 where $\times$ denotes the identification %
$\epsilon_A = -\epsilon_B=\epsilon^+,
\ \gamma\to+0;\ \epsilon_D =-\epsilon_C=\epsilon $. %
 In Ref.~\cite{heusler2007}, the identifications with
the symbols $||$ and $\times$ were called ``column-wise'' and
``crosswise'', respectively.
\section{Consecutive Approximation Orders in Terms
of  Periodic Orbits}
\subsection{Diagonal Approximation}
Let us first consider the diagonal approximation, in which
different orbits are assumed to be uncorrelated, except for the
orbits obtained from each other by time reversal. We express
Eq.(\ref{pseudo_orbits_z1}) in terms of the periodic orbits, not
using pseudo orbits. On account of Eq.(\ref{a_expression}), the
generating function $Z^{(1)}$ can be written in terms of the
periodic orbits:
 \beq Z^{(1)} &=& e^{i(\epsilon_A -
\epsilon_B - \epsilon_C + \epsilon_D )/2 } \left\langle \exp
\left[ \sum_{a} e^{{i\over \hbar } \{ S_a (E) + \Theta_a \} }
f_{AC}^a + e^{-{i\over \hbar } \{ S_a (E) + \Theta_a \} }
{f^a_{BD}}^{\ast} \right]
\right\rangle  , \nonumber \\
f_{AC}^a &=& F_a (E)
\left(  e^{i {T_a \over T_H } \epsilon_A }
-
e^{i {T_a \over T_H } \epsilon_C }   \right) ,  \label{diag1}\\
{f^a_{BD}}^{\ast} &=& F_a^{\ast} (E) \left( e^{-i {T_a \over T_H }
\epsilon_B } - e^{-i {T_a \over T_H } \epsilon_D } \right) .
\nonumber \eeq%
 It can be rewritten as an averaged product over the {\it pairs} $(a,a^{TR})$
  of the mutually time reversed orbits; considering that a time-reversed
orbit has a different sign in the gauge-potential term, we obtain%
 \beq Z^{(1)} &=& e^{i(\epsilon_A -
\epsilon_B - \epsilon_C + \epsilon_D )/2 } \left\langle
\prod_{\left(a,a^{TR}\right) } \exp \left[ e^{{i\over \hbar } S_a
}f_{AC}^a ( e^{{i\over \hbar } \Theta_a } + e^{-{i\over \hbar }
\Theta_a }
) \right. \right. \nonumber \\
& + & \left. \left. e^{-{i\over \hbar } S_a }{f_{BD}^a}^{\ast} (
e^{{i\over \hbar } \Theta_a } + e^{-{i\over \hbar } \Theta_a } )
\right] \right\rangle.
\eeq%
Now let us calculate the average. As the factors in the product
    are assumed to be uncorrelated, the averaged product of the
    factors can be replaced by the product of the averaged factors.
The average over the energy $E$ is replaced by an average
    \begin{equation}
    \langle ... \rangle = \frac{1}{2 \pi \hbar}
    \int_0^{2 \pi \hbar} (...) dS_a
    \end{equation}
over the classical action $S_a$. The terms containing the magnetic
action are averaged over the Gaussian white noise (see
Eq.(\ref{gaussian_weight}) as
\beq \langle\langle e^{2i\Theta_a \over \hbar }\rangle \rangle &=&
e^{-b T/T_H }, \eeq%
where $b=4B^2D T_H /\hbar^2 $. Considering that the
coefficients $f^a_{AC},f^a_{BD}$  are exponentially small for the
relevant long orbits, we expand each factor  up to the second
order and doubly average the result.
Then we find
\beq \left\langle  \exp \left[ e^{{i\over \hbar } S_a }f_{AC}^a (
e^{{i\over \hbar } \Theta_a } + e^{-{i\over \hbar } \Theta_a } )
\right.   +  \left. e^{-{i\over \hbar } S_a }{f_{BD}^a}^{\ast} (
e^{{i\over \hbar } \Theta_a } + e^{-{i\over \hbar } \Theta_a } )
\right] \right\rangle\nonumber\\
\approx 1+f_{AC}^a{f_{BD}^a}^* \left(2+\langle\langle
e^{\frac{i2\Theta}\hbar}\rangle\rangle+\langle\langle
e^{\frac{-i2\Theta}\hbar}\rangle\rangle\right)\nonumber\\
=1+2 f_{AC}^a{f_{BD}^a}^*\left(1+e^{-bT_a/T_H}\right) \nonumber\\
\approx \exp\left[2
f_{AC}^a{f_{BD}^a}^*\left(1+e^{-bT_a/T_H}\right)\right]
\eeq%
It is now straightforward to obtain the expression
 \beq Z_{\rm
diag}^{(1)}  \sim  e^{i(\epsilon_A - \epsilon_B - \epsilon_C +
\epsilon_D )/2 } \nonumber \\\fl  \times  \exp \left[ \sum_a |F_a
(E) |^2 \left( e^{{i T_a \over T_H} \epsilon_A } - e^{{i T_a \over
T_H} \epsilon_C } \right) \left( e^{{-i T_a \over T_H} \epsilon_B
} - e^{{-i T_a \over T_H} \epsilon_D } \right)
\left(1+e^{-bT_a/T_H}\right)\right] ,
\eeq
where summation in the exponent is carried over orbits, not orbit
pairs; for compensation, the factor 2 is dropped.

We can further proceed with the calculation by using Hannay and
Ozorio de Almeida (HOdA) sum rule~\cite{HOdA} \beq \sum_a |F_a|^2
(\cdot ) \sim \int_{T_0}^{\infty } {d T \over T} (\cdot ), \eeq
where the lower limit of the integration is the minimum period
$T_0$; in the semiclassical limit it can be replaced by zero. Then
we immediately obtain the result of the diagonal approximation:
\beq Z_{\rm diag}^{(1)} &=& e^{i(\epsilon_A - \epsilon_B -
\epsilon_C + \epsilon_D)/2}
\nonumber \\
& \times & \left\{
{(\epsilon_C - \epsilon_B ) (\epsilon_A - \epsilon_D  )
\over
(\epsilon_A - \epsilon_B  ) (\epsilon_C - \epsilon_D  )} \right\}
\left\{
{(\epsilon_C - \epsilon_B + i b ) (\epsilon_A - \epsilon_D + i b )
\over
(\epsilon_A - \epsilon_B + i b ) (\epsilon_C - \epsilon_D + i b )}
\right\}.
\eeq

\subsection{Second Order Contribution}
Let us now go beyond the diagonal approximation. In order to
calculate the off-diagonal contributions, we can utilize the
method developed in~\cite{muller,heusler2007}. In that method, the
periodic orbits are divided into links and encounters. An
encounter is a part of an orbit where different links meet
together, and orbit stretches connect the links inside the
encounter region. It is assumed that the contributions come from
the periodic-orbit pairs with the same (or time-reversed) links
which are differently connected inside the encounter regions. Due
to the different connections in the encounter regions, there is a
small but significant difference in actions between the original
and partner orbits. The gauge potential acts on  both links and
encounters. The gauge potential constructively interferes when the
original and partner orbits go in the opposite directions. On the
other hand, when the directions are the same, a destructive
interference takes place.

The off-diagonal contribution
is formally written by using the pseudo-orbit representation
(\ref{pseudo_orbits_z1}) as
\beq
Z^{(1)}_{\rm off} &=&
\sum_{A,B,C,D}\!\!\!\!{}^{'}
\,\,\Bigl\langle
F_A F_B^{\ast} F_C F_D^{\ast} (-1)^{N_C + N_D}
e^{{i\over \hbar} \Delta S }
\Bigr\rangle
\langle \langle e^{{i\over \hbar } \Delta\Theta }\rangle \rangle
\nonumber \\
&&\times e^{{i\over T_H} (T_A \epsilon_A - T_B \epsilon_B + T_C
\epsilon_C - T_D \epsilon_D )} \label{pseudo_orbits_zoff}, \eeq
where $\sum_{A,B,C,D}'$ means a summation over the pseudo orbits
forming links and encounters. $\Delta S$ represents a sum of
action differences in the encounter regions, and $\Delta \Theta$
expresses the effect of interferences of the gauge potential. The
overall sum $Z^{(1)}$ is obtained by multiplying the two
contributions \beq Z^{(1)} &=& Z_{\rm diag}^{(1)} Z_{\rm
off}^{(1)} . \eeq

Let us now calculate the leading (second order) term of the
off-diagonal contribution.  We first discuss the Sieber-Richer
(SR) pairs, one of which is depicted in Figure $1$. A SR pair has
one encounter and two links, and the original and partner orbits
go in the opposite directions in one of the links. In Figure $1$,
the solid curve is the original orbit and the dashed curve is the
partner. A pseudo-orbit index ``A'' or ``C'' is assigned to the
original orbit, and ``B'' or ``D'' is to the partner. There are in
total $16$ possibilities, as listed in Table $1$.
\begin{figure}[ht]
\centering
\includegraphics[width=10.0cm]{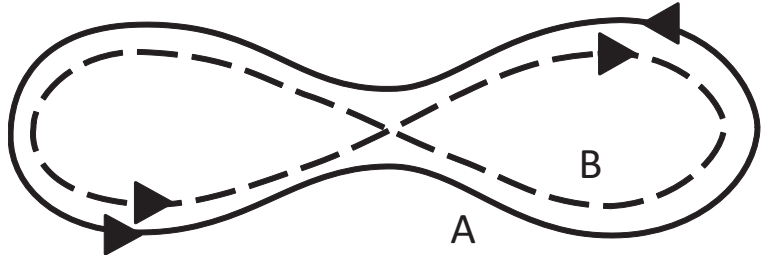}
\caption{An example of the Sieber-Richter (SR)
pairs, where ``A'' is assigned to the original orbit
(the solid curve) and ``B'' is to the partner (the dashed curve).}
\end{figure}
\begin{table}
\centering
\begin{tabular}{|c|c|c|c|c|c|c|c|c|c|c|c|c|c|c|c|c|} \hline
A & $\rsolid$ & $\rsolid$  & $\rsolid$ & $\rsolid$
  & $\lsolid$ & $\lsolid$  & $\lsolid$ & $\lsolid$
  & $\emptyset$      & $\emptyset$       & $\emptyset$      &$\emptyset$
  & $\emptyset$      & $\emptyset$       & $\emptyset$      &$\emptyset$
    \\ \hline
B & $\rdash$  & $\ldash$   & $\emptyset$      & $\emptyset$
  & $\rdash$  & $\ldash$   & $\emptyset$      & $\emptyset$
  & $\rdash$  & $\ldash$   & $\emptyset$      & $\emptyset$
  & $\rdash$  & $\ldash$   & $\emptyset$      & $\emptyset$
\\ \hline
C & $\emptyset$      & $\emptyset$       & $\emptyset$      &$\emptyset$
  & $\emptyset$      & $\emptyset$       & $\emptyset$      &$\emptyset$
  & $\rsolid$ & $\rsolid$  & $\rsolid$ & $\rsolid$
  & $\lsolid$ & $\lsolid$  & $\lsolid$ & $\lsolid$
 \\ \hline
D & $\emptyset$      & $\emptyset$       & $\rdash$  & $\ldash$
  & $\emptyset$      & $\emptyset$       & $\rdash$  & $\ldash$
  & $\emptyset$      & $\emptyset$       & $\rdash$  & $\ldash$
  & $\emptyset$      & $\emptyset$       & $\rdash$  & $\ldash$
\\ \hline
\end{tabular}
\caption{A schematic table for the $16$ possibilities of the
SR pairs. Here $\emptyset$ means the absence in the diagram.}
\end{table}
On the Poincar\'e section within the encounter region, the phase
space can be furnished with canonical coordinates $u$ and $s$,
which are the coordinates in the unstable and stable directions,
respectively. Let us suppose that the original orbit pierces the
Poincar\'e section at the origin $(0,0)$ and at the point $(u,s)$.
To demand that the piercing points in the encounter are mutually
close, we introduce a bound $c$ and assume that $|u|,|s| < c$.
Then we can estimate the duration $t_{\rm enc}$ of the encounter
and the action difference $\Delta S$ of the original and partner
orbits as $ t_{\rm enc} = {1\over \lambda} \ln {c^{2}\over |us|} $
and $\Delta S = us, $ where $\lambda$ is the Lyapunov exponent.
Now we need to estimate the number of encounters in one periodic
orbit of a period $T$. This can be computed as \beq \int_{-c}^c du
ds \int_{0}^{T -2t_{\rm enc}} d t {T \over 2^2 t_{\rm enc}
\Omega}, \label{enb} \eeq where $\Omega^{-1} \left( ={1\over
2\pi\hbar T_H} \right)$ is the ergodic return probability per unit
action. The combinatorial factor $2^2$ implies that one of the two
encounter stretches is chosen as the first stretch and one of the
two directions is chosen as the positive direction. The factor $T
\int_{0}^{T - 2 t_{\rm enc}} {\rm d} t $ indicates that one of the
two piercings occurs in the time interval $[0,T]$ and that the
time $t$ elapsed on one of the links lies in $[0,T-2 t_{\rm
enc}]$. The contribution from the gauge potential has to be taken
into account on one of the links, where the original and partner
orbits go in the opposite directions. In the SR pair, a
constructive interference of the gauge potential does not exist in
the encounter region. Thus the contribution from the terms
depicted in Figure $1$ can be calculated as \beq Z_{{\rm
Fig.1}}^{(1)}&=& \sum_a |F_a|^2 \int ds du \int_{0}^{T -2t_{\rm
enc}} d t {T \over 2^2 t_{\rm enc} \Omega} e^{i su / \hbar} e^{-b
t /T_H } e^{i (\epsilon_A  - \epsilon_B )T/ T_H } . \label{zsr1}
\eeq The HOdA sum rule~\cite{HOdA} changes the sum
 $\sum_a |F_a|^2 T_a $ into $\int dT$. Hence, Eq.(\ref{zsr1}) is
rewritten as \beq {1\over 4} \int ds du \int_{2 t_{\rm
enc}}^{\infty } d T \int_{0}^{T -2t_{\rm enc}} dt {1 \over \Omega
t _{\rm enc}} e^{i su / \hbar} e^{-b t /T_H  } e^{i (\epsilon_A  -
\epsilon_B )T/ T_H } . \eeq

To clarify the contributions from the encounter and links, we
introduce new variables as \beq
T &=& t_1 + t_2 + 2 t_{\rm  enc} ,\\
t &=& t_2  .
\eeq
Then the expression is simplified as
\beq
&&{1\over 4}
\left\{ {1\over T_H } \int_{0}^{\infty} d t_1
e^{i( \epsilon_A - \epsilon_B ) t_1 / T_H  } \right\}
\left\{ {1\over T_H } \int_{0}^{\infty} d t_2
e^{i( \epsilon_A - \epsilon_B ) t_2 / T_H -b t_2 / T_H  } \right\}
\nonumber \\
&\times & T_H^2 \int ds \int du {1 \over \Omega t_{\rm enc}}
e^{isu/\hbar } e^{i (\epsilon_A - \epsilon_B ) 2 t_{\rm enc} / T_H
} . \eeq Now we can identify the contributions from the encounter
and links: in the above expression, the integrals over $t_1$ and
$t_2$ are interpreted as the link-contributions, and the integral
over $(u,s)$ is the encounter-contribution. We expand the
expression in $t_{\rm enc}$ and extract the constant term, which
is expected to survive in the semiclassical limit $\hbar
\rightarrow 0$. Then we obtain \beq
 {1\over 2 i (\epsilon_A - \epsilon_B + ib)} .
\eeq
We repeat similar calculations for the possibilities
listed in Table $1$.  Summing up the results, we
find the contribution from the SR pairs
\beq
Z_{\rm SR}^{(1)} &=&
{2\over i (\epsilon_A - \epsilon_B + ib )}-
{2\over i (\epsilon_A - \epsilon_D + ib )} \nonumber \\
&-&
{2\over i (\epsilon_C - \epsilon_B + ib )}
+
{2\over i (\epsilon_C - \epsilon_D + ib )}  \label{sr_result}
.
\eeq
\begin{figure}[ht]
\centering
\includegraphics[width=10.0cm]{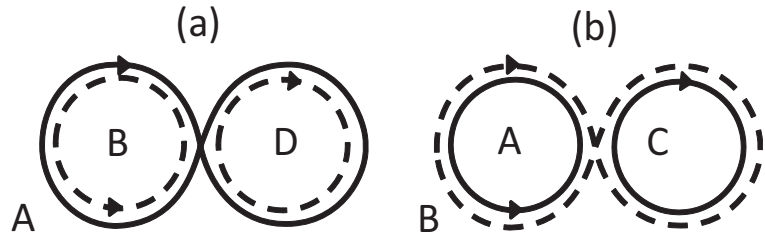}
\caption{Examples of the parallel Sieber-Richter (aSR) pairs.}
\end{figure}

In the evaluation of the second order term, we need to consider
the other types of pseudo-orbit pairs, two examples of which are
shown in Figure $2$. These are called ``parallel Sieber-Richter''
(aSR) pairs, following Ref.~\cite{heusler2007}.  In Figure $2(a)$,
an index ``A'' or ``C'' is assigned to the original orbit (solid
curve), and ``B'' and ``D'' are to the components of the partner
pseudo orbit. In Figure $2(b)$, on the other hand, ``A'' and ``C''
are assigned to the components of the original pseudo orbit, while
``B'' or ``D'' is to the partner orbit. Let us calculate the
contribution from Figure $2(a)$. In both of the two links, the
original orbit and the partner pseudo orbit go in the opposite
directions.  In the encounter, the stretches of the original orbit
and those of the partner pseudo orbit also go in the opposite
directions. Hence, in this case there are constructive
interferences of the gauge potential in all parts of the pair. The
link contribution is given by \beq {1\over T_H}\int_{0}^{\infty} d
t_1 e^{i(\epsilon_A - \epsilon_B )t_1 / T_H -b t_1 /T_H} \times
{1\over T_H}\int_{0}^{\infty} d t_2 e^{i(\epsilon_A - \epsilon_D
)t_2 / T_H -b t_2 /T_H} \eeq and the encounter yields \beq T_H^2
\int du \int ds {1 \over \Omega t_{\rm enc} } e^{i su/\hbar } e^{
i \{ (2\epsilon_A - \epsilon_B -\epsilon_D )/T_H + i4b /T_H \}
t_{\rm enc}} . \eeq In general, when the original (partner) pseudo
orbit passes the encounter $n_+$ and $n_-$ times ($n'_+$  and
$n'_-$ times) in the positive and negative directions,
respectively, an exponential factor ${\rm exp}(i n_{\rm enc}^2 \ b
\ t_{\rm enc}/T_H)$ with \beq n_{\rm enc} = \frac{1}{2} \left| n_+
- n_- - n'_+ + n'_- \right| \eeq appears in the encounter
contribution~\cite{saitonagao2006}. This is a consequence of the
stochastic behavior of the gauge potential
(\ref{gaussian_weight}). Extracting the term surviving in the
semiclassical limit $\hbar \rightarrow \infty$, we obtain the
contribution from Figure $2(a)$ as \beq Z_{{\rm
Fig.}2(a)}^{(1)}&=& {1\over 4} \left\{ {1\over
T_H}\int_{0}^{\infty} d t_1 e^{i(\epsilon_A - \epsilon_B ) t_1 /
T_H
-b t_1 /T_H } \right\}  \nonumber \\
&\times& \left\{
 {1\over T_H}\int_{0}^{\infty} d t_2 e^{i(\epsilon_A - \epsilon_D )t_2 / T_H
-b t_2 /T_H }  \right\} \nonumber \\
&\times&
T_H^2 \int du \int ds {1 \over \Omega t_{\rm enc} } e^{i su/\hbar }
e^{i \{ (2\epsilon_A - \epsilon_B -\epsilon_D )/T_H + i 4 b \}
t_{\rm enc}} \nonumber \\
&=& {-i\over 4} {2 \epsilon_A - \epsilon_B -\epsilon_D + i4b \over
( \epsilon_A - \epsilon_B + ib  )( \epsilon_A - \epsilon_D + ib )}
\eeq A calculation for the diagram in Figure $2(b)$ is similar and
yields %
 \beq Z_{{\rm Fig.}2(b)}^{(1)}&=&{1\over 4} \left\{
{1\over T_H}\int_{0}^{\infty} d t_1 e^{i ( \epsilon_A - \epsilon_B
) t_1 / T_H
-b t_1 /T_H } \right\} \nonumber \\
&\times & \left\{
 {1\over T_H}\int_{0}^{\infty} d t_2 e^{i(\epsilon_C - \epsilon_B )t_2 / T_H
-b t_2 /T_H }  \right\} \nonumber \\
&\times&
T_H^2 \int du \int ds {1 \over \Omega t_{\rm enc} } e^{i su/\hbar }
e^{i \{ (\epsilon_A + \epsilon_C -\epsilon_B )/T_H + i 4 b
\} t_{\rm enc}} \nonumber \\
&=& {-i\over 4} {\epsilon_A + \epsilon_C -\epsilon_B + i4b \over (
\epsilon_A - \epsilon_B + ib  )( \epsilon_C - \epsilon_B + ib  )}
\eeq

In order to obtain the complete set
of the aSR pairs. we need to change
the assignment of the indices $A,B,C$
and $D$ as well as the directions
of the pseudo-orbit components.
There are $64$ diagrams for each of
the types in Figure $2$(a) and Figure $2$(b).
Thus totally $128$ diagrams have to be considered.
Summing up the results by using "Mathematica",
we find the total contribution from the aSR pairs
\beq
Z_{\rm aSR}^{(1)} &=&
2b (\epsilon_{A} - \epsilon_{C} )
(\epsilon_{B} - \epsilon_{D} ) \nonumber \\
&\times&
\left\{
{1\over (\epsilon_C - \epsilon_D + ib)^2
        (\epsilon_A - \epsilon_D + ib)
        (\epsilon_C - \epsilon_B + ib)   }
\right. \nonumber  \\
&-& \left.
{1\over (\epsilon_A - \epsilon_B + ib)^2
        (\epsilon_A - \epsilon_D + ib)
        (\epsilon_C - \epsilon_B + ib)   }
\right\} . \label{asr_result}
\eeq
Note that, in both of the limits $b\to 0$
and $b\to\infty$, this expression vanishes:
It is finite only in the crossover domain.

\subsection{Higher Order Terms}
Let us consider the higher order terms in the periodic-orbit
expansion. Suppose that the pseudo-orbit pair includes $V$
encounters, as schematically drawn in Figure $3$. In the figure,
the ports (the end points of the encounters) are depicted as black
dots. A pseudo orbit is formed by connecting the ports by
stretches within the encounters and by links out of them. In order
to see the general structures of the higher order terms, we need
to count the number of pseudo-orbit pairs, carefully avoiding
overcountings.

As we have seen in the calculation of the SR and aSR pairs, we can
identify the contribution from links and that from the encounters.
That is, a link gives a contribution
\begin{eqnarray}
{i \over
\epsilon_{A {\rm or} C} - \epsilon_{B {\rm or} D} + i\mu b  },
\end{eqnarray}
where $\mu =0$ if the link is traversed with the same sense in
the original and partner pseudo orbits, and $\mu = 1$ if
the senses of traversal are opposite.
On the other hand, the encounter has the
contribution
\begin{eqnarray}
i (m_A \epsilon_A - m_B \epsilon_B +
m_C \epsilon_C - m_D \epsilon_D ) - b n_{\rm enc}^2,
\end{eqnarray}
where $m_A,\cdots,m_D$ are the number of the stretches
belonging to $A,B,C,D$, respectively.

Now we estimate the number of over-countings, when there are $V$
encounters.  Obviously, the ordering of the encounters is
arbitrary. Therefore there are  $V!$ overcountings from the
ordering of the encounters. Suppose that the $j$th encounter has
$\ell_j$ stretches of the original pseudo orbit. Inside the $j$th
encounter, we also overcount $\ell_j$ times, since it is arbitrary
which port is chosen as the first. Since the stretches are no
longer required to point from left to right, we can relabel the
left side as the right side and vice versa. This requires one more
overcounting factor $2^{V}$.

Thus the periodic-orbit expansion of
$Z_{\rm off}^{(1)}$ is formally written as
\beq
Z_{\rm off}^{(1)}
&=& \sum_{V=1}^{\infty} {1\over V! }
\sum_{\ell_1 , \ell_2 , \cdots , \ell_V }
\sum_{{\rm pseudo-orbit \ pairs} } (-1)^{n_C +n_D} \nonumber \\
&& \times {\prod_{j=1}^{V} \left({1\over 2 \ell_{j}} \right)\{
i(m_{A_j} \epsilon_A - m_{B_j} \epsilon_B + m_{C_j} \epsilon_C -
m_{D_j} \epsilon_D ) - b n_{{\rm enc},j}^2 \} \over \prod_{\rm
links} (-i) \{ ( \epsilon_{A {\rm or} C} - \epsilon_{B {\rm or} D}
) + i\mu b
\}}, \nonumber \\
\label{zoff1}
\eeq
where the second sum is over all possible
combinations of the number of stretches, and
the third sum means counting over all the
possible pseudo-orbit pairs when $(\ell_1,
\ell_2, \cdots ,\ell_V )$ are given.

\begin{figure}[ht]
\centering
\includegraphics[width=10.0cm]{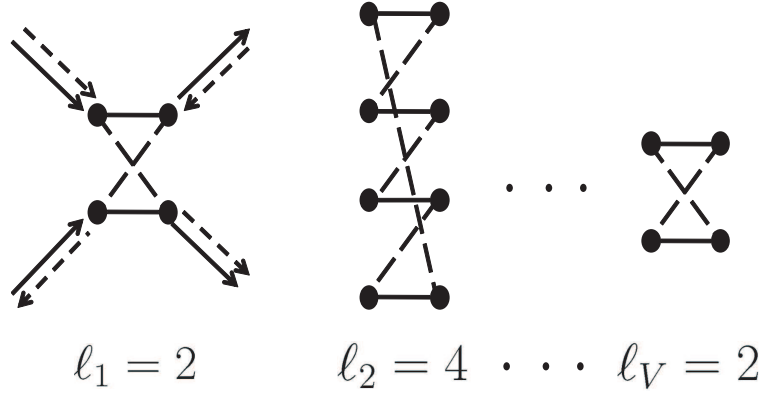}
\caption{A schematic picture of a pseudo orbit pair with $V$
encounters. The links are shown only for the first encounter. The
solid lines depict the original pseudo orbit and the dashed lines
depict the partner.}
\end{figure}

\section{Nonlinear Sigma Model in GOE-GUE Crossover Domain}
Let us examine the random matrix theory prediction for the
generating function and derive an expansion which can directly be
compared with the semiclassical formula. For that purpose, 
using the Bosonic variant of the replica trick we analyze the
nonlinear sigma (NLS) model. Compared with the Fermionic replica
and the supersymmetric version of the sigma model ~\cite{efetov}
the Bosonic replica has the disadvantage  that it yields only the
part $Z^{(1)}$ of the generating function. However we choose it
because it is technically much simpler, and, anyhow, the
complement $Z^{(2)}$ can be obtained from $Z^{(1)}$ by simply
swapping arguments as $\epsilon_C\leftrightarrow \epsilon_D$. As
shown in the following sections, the ensuing perturbative
expansion of the NLS model is equivalent to the semiclassical
periodic-orbit expansion for $Z^{(1)}$.

As derived in Appendix A, the NLS model for the generating
function is written as 
\beq {\cal Z} &=& \lim_{r\to 0} \int d[B] ~ \exp\{ {\cal S}(B) \},
\\
{\cal S}(B )&=&
{i\over 2} (\epsilon_A - \epsilon_B + \epsilon_C - \epsilon_D ) \nonumber \\
&+& {i\over 2} {\rm Tr} (\tilde{\epsilon}_{AC} B B^{\dagger } )
- {i \over 2} {\rm Tr} (\tilde{\epsilon}_{BD}  B^{\dagger } B )
-2s {\rm Tr}(B B^{\dagger}) +2s {\rm Tr} (B\tau_3 B^{\dagger} \tau_3 )
\nonumber \\
&+& {i \over 2} \sum_{m=2}^{\infty}
\left[
{\rm Tr} \left\{
\tilde{\epsilon}_{AC} (B B^{\dagger } )^m \right\} -
{\rm Tr} \left\{ \tilde{\epsilon}_{BD} ( B^{\dagger } B )^m \right\} \right]
-2s \sum_{m=2}^{\infty} {\rm Tr} (B B^{\dagger})^m
\nonumber \\
&-& s\sum_{m=1}^{\infty } \sum_{m'=1}^{\infty } {\rm Tr}
\left\{ (B B^{\dagger} )^m \tau_3 ( B B^{\dagger})^{m'} \tau_3 \right\}
\nonumber \\
&-& s
\sum_{m=1}^{\infty } \sum_{m'=1}^{\infty } {\rm Tr}
\left\{ (B^{\dagger} B)^m \tau_3 ( B^{\dagger} B)^{m'} \tau_3 \right\}
\nonumber \\
&+&2s \!\!
\sum_{m,m'=0 \atop (m,m')\ne (0,0)}^{\infty}
{\rm Tr} \left\{ (B B^{\dagger})^m B \tau_3 (B^{\dagger} B)^{m'} B^{\dagger}
\tau_3 \right\} \, , \label{s_exp}
\eeq
where the measure in the integral is defined as 
$d[B]=\prod_{i,k} d\mbox {Re} B_{ik}\, d \mbox{Im}
B_{ik}$. 
$\tilde{\epsilon}_{AC}$, $\tilde{\epsilon}_{BD}$
and $\tau_3$ are $2r \times 2r$ matrices given by
\beq
\tilde{\epsilon}_{AC} &=&
\left(
\begin{array}{ll}
\hat{\epsilon}_{AC} &  0 \\
0 & \hat{\epsilon}_{AC}
\end{array}
\right) , \qquad
\hat{\epsilon}_{AC} = {\rm diag} (\epsilon_A , \overbrace{\epsilon_C, \cdots, \epsilon_C}^{r-1} ) ,
\label{epsac} \\
\tilde{\epsilon}_{BD} &=&
\left(
\begin{array}{ll}
\hat{\epsilon}_{BD} & 0 \\
0 & \hat{\epsilon}_{BD}
\end{array}
\right) , \qquad
\hat{\epsilon}_{BD} = {\rm diag} (\epsilon_B , \overbrace{\epsilon_D, \cdots, \epsilon_D}^{r-1} ) ,
\label{epsbd} \\
\tau_3 &=& {\rm diag}(\overbrace{1,1,\cdots,1}^{r}, \overbrace{-1,
-1, \cdots, -1}^{r} ) . \label{tau3} \eeq
The integrations are performed over all the matrix elements of $B$
with the range $(-\infty , \infty)$. The variable $r$ is the
number of replica spaces, and we finally take the limit $r\to 0$.

\subsection{Diagonal Term}
Let us calculate the diagonal term ${\cal Z}_{\rm diag}$
of the generating function. The diagonal contribution comes
from the quadratic terms of ${\cal S}(B)$ as
\beq
{\cal Z}_{\rm diag} &=& e^{i(\epsilon_A - \epsilon_B - \epsilon_C + \epsilon_D)/2}
\lim_{r\to 0}\int d[B] \exp \left\{
{i\over 2} {\rm Tr} (\tilde{\epsilon}_{AC} B B^{\dagger} )
-{i\over 2} {\rm Tr} (\tilde{\epsilon}_{BD} B^{\dagger} B )
\right. \nonumber \\
&&
\left. -2s {\rm Tr} (B B^{\dagger} )
+ 2s {\rm Tr} (B \tau_3 B \tau_3 )
\right\} . \label{gaussian}
\eeq
In order to explicitly calculate this integral, we decompose
the $B$ matrix as
\beq
B &=&
\left(
\begin{array}{cc}
B^{++} , & B^{+-} \\
B^{-+} , & B^{--} \\
\end{array}
\right) ,
\eeq
where each matrix $B^{\alpha \beta}$ is an
$r\times r$ matrix.  We note that the symmetry
(\ref{symmetry2}) requires the relations
\beq
(B^{++})^{\dagger} &=& - (B^{--})^{T}, \\
(B^{-+})^{\dagger} &=& - (B^{+-})^{T}.
\eeq
Using these relations, we calculate (\ref{gaussian}) as
\beq
{\cal Z}_{\rm diag} &=& e^{i(\epsilon_A - \epsilon_B - \epsilon_C + \epsilon_D)/2}
\nonumber \\ & \times &
\lim_{r\to 0}\! \int d[B] \exp \left[
\sum_{\alpha,\beta =\pm}\sum_{k,\ell=1}^{r}
\left\{ {i\over 2} (\hat{\epsilon}_{AC,k} - \hat{\epsilon}_{BD,\ell} ) -2s +2s \alpha \beta
\right\} |B_{k,\ell}^{\alpha , \beta }|^2 \right] \nonumber \\
&=& e^{i(\epsilon_A - \epsilon_B - \epsilon_C + \epsilon_D)/2}
\nonumber \\
& \times & \lim_{r\to 0}
{-\pi \over i (\epsilon_{A} - \epsilon_{B} ) -8s}
~{-\pi \over i (\epsilon_{A} - \epsilon_{B} ) }
\left\{ {-\pi \over i (\epsilon_{A} - \epsilon_{D} ) -8s}
~{-\pi \over i (\epsilon_{A} - \epsilon_{D} ) } \right\}^{r-1} \nonumber \\
&\times&\!\!
\left\{ {-\pi \over i (\epsilon_{C} - \epsilon_{B} ) -8s}
{-\pi \over i (\epsilon_{C} - \epsilon_{B} ) } \right\}^{r-1} \!\!
\left\{ {-\pi \over i (\epsilon_{C} - \epsilon_{D} ) -8s}
{-\pi \over i (\epsilon_{C} - \epsilon_{D} ) }
\right\}^{(r-1)(r-1)}. \nonumber \\
\eeq
Taking the limit $r\to 0$, we obtain
\beq
{\cal Z}_{\rm diag}
&=& e^{i(\epsilon_A - \epsilon_B - \epsilon_C + \epsilon_D)/2}
\nonumber \\
& \times & \left\{
{
(\epsilon_A - \epsilon_D  + i8s ) (\epsilon_C - \epsilon_B + i 8s)
\over
(\epsilon_A - \epsilon_B  + i8s ) (\epsilon_C - \epsilon_D + i 8s)
}
\right\}
\left\{
{
(\epsilon_A - \epsilon_D ) (\epsilon_C - \epsilon_B )
\over
(\epsilon_A - \epsilon_B ) (\epsilon_C - \epsilon_D )
}
\right\}.
\eeq
This is identical to the semiclassical formula, if $8s$ is
replaced by $b$.

\subsection{Expansion of the Generating Function}
The generating function ${\cal Z}$ is formally
expanded as
\beq
{\cal Z}={\cal Z}_{\rm  diag} (1 + {\cal Z}_{\rm off}),
\eeq
\beq
{\cal Z}_{\rm off} & = & \lim_{r\to 0}
\left\langle \exp \left( \sum_{\ell=2}^{\infty }
{\cal F}_{\ell} \right)\right\rangle  -1 \nonumber \\
& = & \lim_{r\to 0} \left\{ \sum_{\ell }\langle {\cal F}_{\ell} \rangle
+  {1\over 2!} \sum_{\ell , \ell' }
\langle {\cal F}_{\ell}{\cal F}_{\ell '}
 \rangle + \cdots \right\} \! ,
\label{zformaloff}
\eeq
where ${\cal F}_{\ell}$ is written as
\beq
{\cal F}_{\ell} &=&
 {i \over 2}
{\rm Tr} \{ \tilde{\epsilon}_{AC} (B B^{\dagger } )^{\ell} \}
-
 {i \over 2} {\rm Tr} \{
\tilde{\epsilon}_{BD} ( B^{\dagger } B )^{\ell} \}
-2s {\rm Tr} (B B^{\dagger})^{\ell}
\nonumber \\
&-&s\sum_{m=1}^{\ell -1 } {\rm Tr}
\left\{ (B B^{\dagger} )^m \tau_3 ( B B^{\dagger})^{\ell -m} \tau_3 \right\}
- s \sum_{m=1}^{\ell -1} {\rm Tr}
\left\{ (B^{\dagger} B)^m \tau_3 ( B^{\dagger} B)^{\ell -m }  \tau_3 \right\}
\nonumber \\
&+& 2s\sum_{m= 0}^{\ell -1}
{\rm Tr} \{ (B B^{\dagger})^m B \tau_3 (B^{\dagger} B)^{\ell - m -1}
B^{\dagger} \tau_3 \}.
\eeq
The symbol $\langle ... \rangle$ in Eq.(\ref{zformaloff}) means
an average computed over the Gaussian weight in (\ref{gaussian})
with a normalization factor. Perturbative expansions will be
derived by means of the Wick theorem on this Gaussian average.
Thus the Gaussian average plays a crucial role in establishing
the relationship between the periodic-orbit theory and
random matrices. The Gaussian averages of the two-point
correlation functions of $B$ are evaluated as
\beq
\left\langle \,  ( B_{k ,  \ell}^{  \alpha   \beta} )^{\ast}  \,
           B_{k' , \ell '}^{\alpha ' \beta ' } \,  \right\rangle
 &=&-{ \delta_{k, k'} \delta_{\ell , \ell '}
\delta_{\alpha , \alpha '}\delta_{\beta , \beta '}
\over i (\hat{\epsilon}_{AC, k} - \hat{\epsilon}_{BD, \ell})
- 4s |\alpha - \beta|} ,
\label{w1} \\
\left\langle \,   B_{k , \ell}^{\alpha \beta} \,
           B_{k' , \ell '}^{\alpha ' \beta '} \, \right\rangle
 &=&{ \delta_{k, k'} \delta_{\ell , \ell '}
\delta_{\alpha , \bar{\alpha} '}\delta_{\beta , \bar{\beta} '}
\over i (\hat{\epsilon}_{AC, k} - \hat{\epsilon}_{BD, \ell}) - 4s |\alpha - \beta|}
\label{w2} ,
\eeq
where $\bar{\alpha} '$ and $\bar{\beta} '$ denote the opposites of
the signs $\alpha '$ and $\beta '$, respectively, and
$k,k',\ell$ and $\ell '$ take the values $1,\cdots, r$.

In order to obtain a simple expression for ${\cal F}_{\ell}$ in
terms of the matrix elements of $B$, we make use of the
cyclic property of a trace, i.e.,
${\rm Tr}({\cal O}_1 {\cal O}_2 \cdots {\cal O}_L)
= {\rm Tr}({\cal O}_2 {\cal O}_3 \cdots {\cal O}_{1})
= \cdots = {\rm Tr}({\cal O}_L {\cal O}_1 \cdots {\cal O}_{L-1})
$. Due to this cyclic property, $\ell$ equivalent but
seemingly different expressions for ${\cal F}_{\ell}$
are obtained. Taking the average of these expressions,
we find the following compact formula
\beq
{\cal F}_{\ell }
&=&
{1\over \ell}
\sum_{\alpha_1,\cdots ,\alpha_{2\ell}}
\sum_{k_1,\cdots , k_{2\ell}}
B_{k_1 , k_2}^{\alpha_1 , \alpha_2 }
( B_{k_3 , k_2}^{\alpha_3 , \alpha_2 } )^{\ast}
\cdots
B_{k_{2\ell -1} , k_{2\ell -1}}^{\alpha_{2\ell -1} , \alpha_{2\ell} }
( B_{k_1 , k_{2\ell }}^{\alpha_1 , \alpha_{2\ell} } )^{\ast}  \nonumber \\
&\times& {1 \over 2} \left[ \sum_{j=1}^{\ell}
i(\hat{\epsilon}_{AC,k_{2j-1} } - \hat{\epsilon}_{BD,k_{2j} } ) -8
s \left\{ {1 \over 2} \sum_{j=1}^{\ell} ( \alpha_{2 j } -
\alpha_{2j -1} ) \right\}^2 ~~\right] . \label{fell} \eeq As
clarified below, the squared factor corresponds to the factor
$n_{\rm enc}^2$ in the semiclassical
expression~\cite{saitonagao2006}. In order to evaluate the
expansion (\ref{zformaloff}), general correlation functions of $B$
are necessary. We are able to calculate them by using the Wick
theorem. In the next subsection, we obtain the second order term
which yields the leading off-diagonal contribution.

\subsubsection{Second Order Contribution}
Second order contribution in ${\cal Z}_{\rm off}$
results from the quartic terms in $B$. The quartic
terms are simply given by
\beq
\langle {\cal F}_2 \rangle
&=&\sum_{\alpha_1 , \cdots , \alpha_4 =\pm}~
\sum_{k_1 , \cdots , k_4=1}^{r} \left( {1\over 2}\right)^2
\nonumber \\
&\times&
\left\{
i[\hat{\epsilon}_{AC,k_1}+\hat{\epsilon}_{AC,k_3}
-
(\hat{\epsilon}_{BD,k_2}+ \hat{\epsilon}_{BD,k_4} )]
-8s \left( {\alpha_1 + \alpha_3 \over  2} -
{ \alpha_2 + \alpha_4 \over 2} \right)^2
\right\} \nonumber \\
&\times&
\Bigl \langle \,
B_{k_1 , k_2}^{\alpha_1 , \alpha_2} \,
( B_{k_3 , k_2}^{\alpha_3 , \alpha_2} )^{\ast} \,
B_{k_3 , k_4}^{\alpha_3 , \alpha_4} \,
( B_{k_1 , k_4}^{\alpha_1 , \alpha_4} )^{\ast} \,
\Bigr\rangle. \label{2nd}
\eeq
We employ the Wick theorem to write this in terms of
the two-point correlation functions (\ref{w1}) and
(\ref{w2}). That is, a four-point correlation function
is expanded as
\beq
\Bigl \langle \,
B_{k_1 , k_2}^{\alpha_1 , \alpha_2} \,
( B_{k_3 , k_2}^{\alpha_3 , \alpha_2} )^{\ast} \,
B_{k_3 , k_4}^{\alpha_3 , \alpha_4} \,
( B_{k_1 , k_4}^{\alpha_1 , \alpha_4} )^{\ast} \,
\Bigr\rangle
&=&
\Bigl \langle
B_{k_1 , k_2}^{\alpha_1 , \alpha_2}
B_{k_3 , k_4}^{\alpha_3 , \alpha_4}
 \Bigr\rangle
\Bigl\langle
( B_{k_3 , k_2}^{\alpha_3 , \alpha_2} )^{\ast}
( B_{k_1 , k_4}^{\alpha_1 , \alpha_4} )^{\ast}
\Bigr\rangle  \nonumber \\
&+&
\Bigl \langle
B_{k_1 , k_2}^{\alpha_1 , \alpha_2}
( B_{k_3 , k_2}^{\alpha_3 , \alpha_2} )^{\ast}
\Bigr\rangle
\Bigl\langle
B_{k_3 , k_4}^{\alpha_3 , \alpha_4}
( B_{k_1 , k_4}^{\alpha_1 , \alpha_4} )^{\ast}
\Bigr\rangle \nonumber \\
&+&
\Bigl \langle
B_{k_1 , k_2}^{\alpha_1 , \alpha_2}
( B_{k_1 , k_4}^{\alpha_1 , \alpha_4} )^{\ast}
\Bigr\rangle
\Bigl\langle
B_{k_3 , k_2}^{\alpha_3 , \alpha_2}
( B_{k_3 , k_4}^{\alpha_3 , \alpha_4} )^{\ast}
\Bigr\rangle . \nonumber
\eeq
In the periodic-orbit theory, the SR and aSR pairs
give the corresponding contribution to these
terms. Let us consider the contribution from the
first term
\beq
&&\sum_{\alpha_1 , \alpha_2 }
\sum_{k_1 ,k_2}\left( {1\over 2}\right)^2
i( 2 \hat{\epsilon}_{AC,k_1} - 2 \hat{\epsilon}_{BD,k_2}) \nonumber \\
&\times&\left\{ {1 \over i (\hat{\epsilon}_{AC, k_1} -
\hat{\epsilon}_{BD, k_2}) - 4s |\alpha_1 - \alpha_2|}
\right\}
\left\{ {1 \over i (\hat{\epsilon}_{AC, k_1}
- \hat{\epsilon}_{BD, k_2}) - 4s |\alpha_1 + \alpha_2|}
\right\} .
 \nonumber \\
\label{sr11}
\eeq
This expression have $16$ terms after taking the replica limit
$r\to 0$, which is consistent with the the number of the SR pairs
in the semiclassical theory. As easily checked, this reproduces
the result in (\ref{sr_result}), if $8s$ is replaced by $b$.
The second term generates the contribution corresponding to that
of the aSR pairs of the type in Figure $2(a)$:
\beq
&&
\sum_{\alpha_1 , \alpha_2 , \alpha_4}
\sum_{k_1 , k_2, k_4}
\left( {1\over 2}\right)^2 \nonumber \\
&\times& \left[
i \{ 2 \hat{\epsilon}_{AC,k_1}-
(\hat{\epsilon}_{BD,k_2}+ \hat{\epsilon}_{BD,k_4} ) \}
-8s \left(\alpha_1 -  {\alpha_2 + \alpha_4 \over 2}\right)^2
\right] \nonumber \\
&\times&
\left\{ {1\over i(\hat{\epsilon}_{AC ,k_1} - \hat{\epsilon}_{BD , k_2} )-8s
\left|{ \alpha_1 - \alpha_2 \over 2}\right|} \right\}
\left\{
{1\over i(\hat{\epsilon}_{AC ,k_1} - \hat{\epsilon}_{BD , k_4}) -8s
\left|{\alpha_1 - \alpha_4 \over 2}\right|} \right\}. \nonumber \\
\label{asr11}
\eeq
This has $64$ terms after taking $r\to 0$. Finally the third term
gives the contribution corresponding to that of the aSR pairs
of the type in Figure $2(b)$:
\beq
&&
\sum_{\alpha_1 , \alpha_3 , \alpha_2}
\sum_{k_1 , k_3, k_2}
\left( {1\over 2}\right)^2 \nonumber \\
&\times & \left\{
i(\hat{\epsilon}_{AC,k_1}+
\hat{\epsilon}_{AC,k_3}- 2 \hat{\epsilon}_{BD,k_2})
-8s \left(\, {\alpha_{1} + \alpha_3 \over 2} - \alpha_{2}\right)^2
\right\} \nonumber \\
&\times&
\left\{ {1\over i(\hat{\epsilon}_{AC ,k_1} -  \hat{\epsilon}_{BD , k_2} )-8s
\left|{ \alpha_1 - \alpha_2 \over 2 }\right|} \right\}
\left\{ {1\over i(\hat{\epsilon}_{AC ,k_3} -  \hat{\epsilon}_{BD , k_2}) -8s
\left|{\alpha_3 - \alpha_2 \over 2}\right|} \right\}. \nonumber
\\ \label{asr22}
\eeq
Eqs. (\ref{asr11}) and (\ref{asr22}) totally yield $128$ terms,
and reproduce the result in (\ref{asr_result}).

\subsection{Relation between Semiclassical Formulas and Random Matrix Results}
Let us consider how the
semiclassical formulas are related to random matrix results.
In the expansion (\ref{zformaloff}), the factor ${\cal F}_{\ell}$
      corresponds to the $\ell$-encounter, while the $V$-th
      order terms in ${\cal F}_{\ell}$ are related to the terms
      with $V$ encounters.
The factors $B$ and $B^{\ast}$ represent the ports of the
encounters. The contraction lines in the Wick theorem correspond
to the links. Let us consider the factors
$B_{k_i,k_{i+1}}^{\alpha_i , \alpha_{i+1}}$ and
$(B_{k_{j+1},k_{j}}^{\alpha_{j+1} , \alpha_{j} } )^{\ast}$ in
${\cal F}_{\ell}$ and the corresponding ports. Note that
$\alpha_{2 l + 1}$ and $k_{2 l + 1}$ are identified with
$\alpha_1$ and $k_1$, respectively. The ports are connected inside
the $\ell$-encounter as follows. If $i+1 = j$, the original pseudo
orbit connects the ports. On the other hand, if $i=j+1$, the
partner connects them. The superindex $\alpha$ on the left (right)
side specifies the direction of the partner (original) pseudo
orbit. The subindex $k$ on the left (right) side determines the
assignment of the index $B$ or $D$ ($A$ or $C$). Using these
rules, we can easily interpret Eq.(\ref{sr11})-(\ref{asr22}) as
the SR and aSR pairs.

As an example, let us examine the following case \[
\langle
\Wwick
{22}{
 <1 { B^{\alpha_1  \alpha_2  }_{k_1  k_2 } }
 >1 {(B^{\alpha_3 \alpha_2 }_{k_3  k_2 })^{\ast}}
 <2 { B^{\alpha_3  \alpha_4 }_{k_3  k_4 }  }
    {(B^{\alpha_1  \alpha_4 }_{k_1  k_4  })^{\ast}}
    { B^{\alpha_1 '\alpha_2 '}_{k_1 ' k_2 ' }  }
    {(B^{\alpha_3 ' \alpha_2' }_{k_3 '  k_2 ' })^{\ast} }
    { B^{\alpha_3 '\alpha_4 '}_{k_3 ' k_4 ' }  }
    { (B^{\alpha_5 ' \alpha_4' }_{k_5'  k_4 ' })^{\ast} }
 >2 { B^{\alpha_5 '\alpha_6 '}_{k_5 ' k_6 ' }  }
    { (B^{\alpha_1 ' \alpha_6 '}_{k_1 '  k_6 '})^{\ast}  }
}
{323}{
    { B^{\alpha_1  \alpha_2  }_{k_1  k_2 } }
    {(B^{\alpha_3 \alpha_2 }_{k_3  k_2 })^{\ast}}
    { B^{\alpha_3  \alpha_4 }_{k_3  k_4 }  }
  <1{(B^{\alpha_1  \alpha_4 }_{k_1  k_4  })^{\ast}}
  <2{ B^{\alpha_1 '\alpha_2 '}_{k_1'  k_2 ' }  }
  >2{(B^{\alpha_3 ' \alpha_2' }_{k_3'  k_2 ' })^{\ast} }
  >1{ B^{\alpha_3 '\alpha_4 '}_{k_3 ' k_4 ' }  }
  <3{ (B^{\alpha_5 ' \alpha_4' }_{k_5'  k_4 ' })^{\ast} }
    { B^{\alpha_5 '\alpha_6 '}_{k_5 ' k_6 ' }  }
  >3{ (B^{\alpha_1 ' \alpha_6 '}_{k_1'  k_6 '})^{\ast}  }
}
\rangle    ,
\]
where the lines imply contractions. This example consists of two
encounters with $2$ and $3$ stretches. In Figure $4$, the solid
curves depict the original pseudo orbit, and the dashed curves
depict the partner. The port $B$ is located on the left side and
$B^{\ast}$ is on the right side of each encounter. In the figure,
only the port indices and the connections among them are written.
The thin curves (lines) express the connections inside the
encounters, and the bold curves express the links generated by
taking the contractions. Inside each encounter, the ports are
connected by following the explained rules. The directions of the
links and the encounter stretches are also shown in the figure .
If the link connects $B$ and $B^{\ast}$, then their left
superindices (e.g., $\alpha_1$ and $\alpha_3'$) are the same and
so are their right superindices (e.g., $\alpha_4$ and
$\alpha_4'$). On the other hand, if the link connects $B$ and $B$,
or $B^{\ast}$ and $B^{\ast}$, their left (right) superindices
become the opposite. These properties originate from (\ref{w1})
and (\ref{w2}). The signs $\alpha$ can consistently be interpreted
as the directions of the links. In Figure 4, the signs $\alpha$
are $\alpha_1=\alpha_3=\alpha_3'=\alpha_1' = - \alpha_5'=+$,
$\alpha_4=-\alpha_6 ' = -\alpha_4 ' = -$, $\alpha_2'=-$, and
$\alpha_2=+$.

In the semiclassical theory, for each topological
structure of the pseudo-orbit pairs, we have to
consider all the possible ways of assigning either
$A$ or $C$ to the components the original pseudo orbit,
and assigning either $B$ or $D$ to those of the partner.
We see that each component accompanies a
factor $-1$, if $C$ or $D$ is assigned to it.
In the replica treatment of random matrices, on
the other hand, if $C$ or $D$ is assigned,
the summation over all the possible choices of the indices
leads to a factor $r-1$, which goes to $-1$
in the limit $r \rightarrow 0$. Hence the sign finally
becomes the same as that in the semiclassical formula.
\begin{figure}[ht]
\centering
\includegraphics[width=10.0cm]{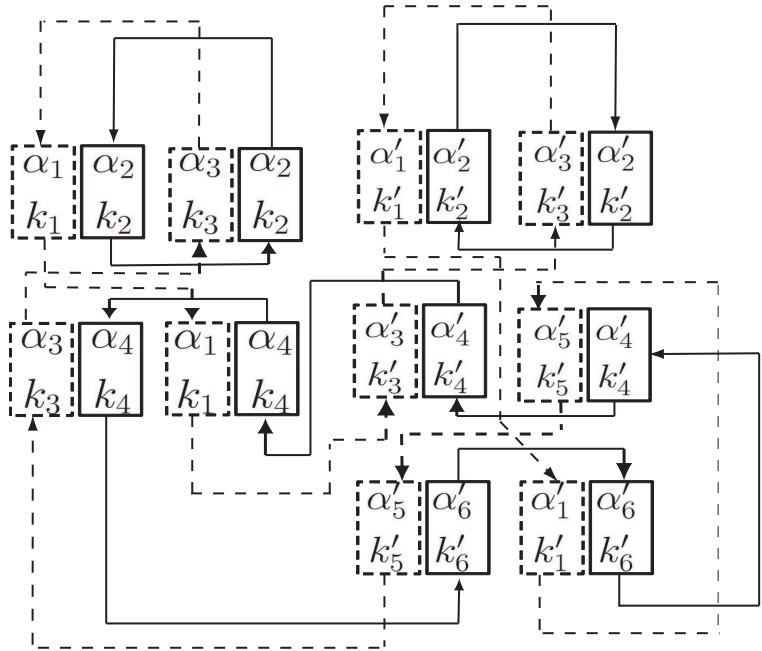}
\caption{An example which consists of two encounters. Solid curves
depict the original pseudo orbit, while the dashed curves depict
the partner. The links correspond to the contractions in the Wick
theorem. The directions of the links and the encounter stretches
are also shown. In this case,
 $\alpha_1=\alpha_3=\alpha_3'=\alpha_1' = - \alpha_5'=+$,
$\alpha_4=-\alpha_6 ' = -\alpha_4 ' = -$, $\alpha_2'=-$, and $\alpha_2=+$.}
\end{figure}

Finally, we discuss how many times the same structure appears,
which gives the overcounting factor. Let us consider the
$\ell$-encounter term Eq.(\ref{fell}), and suppose that the
contractions are already given.

First we change the indices as
\beq
(\alpha_{2m-1} , k_{2m-1}) &\to & (\alpha_{2m-1 + p} , k_{2m-1 + p}) , \\
(\alpha_{2m} , k_{2m}) &\to & (\alpha_{2m + p} , k_{2m + p})
\eeq
for $m=1,\cdots , \ell$. For each value of $p=1,\cdots ,\ell$, we
find the same pseudo orbit pairs. This is related to the fact that
we can arbitrarily choose one of the $\ell$ ports as the first.

Next we exchange the left and right ports
in addition to the transformation
\beq
(\alpha_1 , k_1 ) & \to  &(\bar{\alpha}_3 , k_3 ), \\
(\alpha_3 , k_3 ) & \to  &(\bar{\alpha}_1 , k_1 ), \\
(\alpha_{m+3} , k_{m+3} ) & \to  &(\bar{\alpha}_{2\ell -1 + m} , k_{2\ell + 1 -m } ), \quad m\ge 1 ,\\
(\alpha_{2\ell +1 -m} , k_{2\ell +1 -m} ) & \to
&(\bar{\alpha}_{3+m} , k_{3+m} ), \quad m\ge 1 . \eeq That is,
changing the layout in Figure 4, we put the indices of $B$ and
$B^{\ast}$ on the right and left sides, respectively. Then we
recover the original pair of the pseudo orbits. This procedure
yields two identical  pairs. This related to the fact that we
can relabel the left side of an encounter as the right side and
vice versa in the semiclassical argument.

Thus we found that the overcounting factor was
in total $2 \ell$ for each structure of the
pseudo-orbit pairs.

\subsection{Diagrammatic Expansion of the Generating Function}
Now we are able to write down a diagrammatic expansion of
${\cal Z}_{\rm off}$. Using (\ref{zformaloff}) and the Wick
theorem, we obtain a formal expression
\beq
{\cal Z}_{\rm off} &=&
\sum_{V=1}^{\infty} {1\over V !}
\sum_{\ell_1, \ell_2,\cdots , \ell_V}
\Bigl\langle
{\cal F}_{\ell_1} {\cal F}_{\ell_2} {\cal F}_{\ell_3} \cdots
{\cal F}_{\ell_V}
\Bigr\rangle \nonumber \\
&=&
\lim_{r \rightarrow 0}
\sum_{V =1}^{\infty} {1\over V! }
\sum_{\ell_1, \ell_2,\cdots , \ell_V}
\sum_{\rm contractions} \sum_{k,\alpha}^{}{}^{'} \nonumber \\
&\times& \prod_{j=1}^{V}  {({1/ 2 \ell_{j}})\{ i(m_{A_j}
\epsilon_{A} - m_{B_j} \epsilon_{B} + m_{C_j} \epsilon_{C} -
m_{D_j} \epsilon_{D} ) - 8s \, n_{{\rm enc},j}^2 \} \over (-1) \{
i (\epsilon_{A {\rm or} C} - \epsilon_{B {\rm or} D} ) - 8s \, \mu
\}
} , \nonumber \\
\eeq where the symbol $\sum_{k,\alpha}^{}{}^{'} $ stands for the
sum over all allowed $k$ and $\alpha$ after taking the
contractions. The corresponding values of $m_{A_j},\cdots,
m_{D_j}$, $n_{{\rm enc}, j}$, $\mu$ and the assignments of
$\epsilon_{A {\rm or} C}$, $\epsilon_{B {\rm or} D}$ are
diagrammatically determined for each term. Note that a factor $1/2
\ell_{j}$ is included in order to make up for the overcountings
discussed above. We conjecture that this diagrammatic expansion is
identified with the periodic-orbit expansion (\ref{zoff1}) derived
in \S 3.

\section{Summary}
For a classically chaotic system in a weak magnetic field, the
generating function of the spectral correlation was studied by
using the periodic-orbit theory. The diagonal and leading
off-diagonal terms in the periodic-orbit expansion were explicitly
calculated. Both the non-oscillatory and oscillatory terms were in
agreement with the prediction of the random matrix theory.
Moreover we clarified the general structures of the higher order
terms. In order to see the correspondence to the result of the
random matrix theory, we analyzed the nonlinear sigma (NLS) model
in the crossover domain between the GOE and GUE classes. It was
conjectured that the diagrammatic expansion of the NLS model could
be identified with the periodic-orbit expansion.

\section*{Acknowledgements}
The authors are grateful to Prof. Fritz Haake and Dr. Stefan Heusler 
for stimulating discussion and their continuous interest. 
They also thank Prof. Martin
Zirnbauer for providing his original result\cite{Zirnbauer} before
publication. 
This work is partially supported by Japan Society for the
Promotion of Science (KAKENHI 20540372) and by the
Sonderforschungsbereich TR 12 of the Deutsche
Forschungsgemeinschaft.

\appendix
\section{Random Matrix Theory in GOE-GUE Crossover Domain}
\subsection{Derivation of the Nonlinear Sigma Model}
In this Appendix, we derive the nonlinear sigma model in the
crossover domain between the GOE and GUE classes of random
matrices. The probability distribution of the random
matrix $H$ reads
\beq
P (H) dH & =  &
{\cal N} \exp\left[ - {N \over 4
\left\{ (\sigma_{\rm O} - \sigma_{\rm U} )
e^{-2\tau } + \sigma_{\rm U} \right\} }
\left\{ \sum_{j=1}^{N} H_{jj}^2  + 2 \sum_{j < \ell } ({\rm Re} H_{j \ell })^2
\right\} \right. \nonumber \\
&& \left. - {N \over 2\sigma_{\rm U} \left( 1 - e^{-2\tau } \right) }
\sum_{j < \ell } ({\rm Im} H_{j\ell })^2
\right] dH , \label{pofH}
\eeq
where ${\cal N}$ is a normalization factor. Throughout this Appendix, we use
the symbol ${\cal N}$ for normalization factors, even if their values are
different. The distribution (\ref{pofH}) reproduces
the GOE distribution $P_{\rm GOE}$ in the limit
$\tau\to 0$, and the GUE distribution
$P_{\rm GUE}$ in the limit $\tau\to \infty$, respectively:
\beq
P(H) &\to &P_{\rm GOE } (H) \nonumber \\
&\propto &
\exp
\left[
-{N\over 4 \sigma_{\rm O}}
\left\{ \sum_{j=1}^{N} H_{jj}^2 + 2 \sum_{j < \ell }
({\rm Re} H_{j \ell})^2
\right\}
\right]
\,
 \prod_{j < \ell} \delta ( {\rm Im} H_{j\ell }) ,   \quad \tau \to 0 ~,
\nonumber \\
P(H) & \to & P_{\rm GUE } (H)  \nonumber \\
& \propto &
\exp
\left[
-{N\over 4 \sigma_{\rm U}}
\left\{ \sum_{j=1}^{N} H_{jj}^2 + 2 \sum_{j < \ell }
\left( ({\rm Re} H_{j \ell})^2 + ({\rm Im} H_{j \ell})^2
\right)
\right\}
\right] , ~\tau\to\infty ~. \nonumber
\eeq
We use the Bosonic replica trick to calculate the generating function 
defined by (1) and derive the nonlinear sigma model (NLS). 
Here we use the notation
${\cal Z}$ to discriminate the NLS expression from 
the random matrix expression (3).
The generating function 
 is formally written
in terms of the vectors $\psi_A , \psi_{B} , \psi_{C_k}$ and $\psi_{D_k}$
($k=1,\cdots, r-1$), each of which consists of $N$ elements:
\begin{eqnarray}
{\cal Z} 
&=& \lim_{r\to 0 }
\int d[\psi] e^{i \psi^{\dagger} L {\bm E} \psi} \nonumber  \\
&\times& \left\langle
\exp \left(
-i \psi_A^{\dagger}  H \psi_{A}
-i\sum_{k=1}^{r-1} \psi_{C_k}^{\dagger} H \psi_{C_k}
+i \psi_B^{\dagger} H \psi_{B}
+i\sum_{k=1}^{r-1} \psi_{D_k}^{\dagger} H \psi_{D_k}
\right)
\right\rangle  , \nonumber \\
\end{eqnarray}
where $\psi$ is a vector composed of $\psi_A , \psi_{C_k} \, (k=1,\cdots ,r-1),
\psi_{B}$, and $\psi_{D_k} \, (k=1, \cdots ,r-1)$. The symbol 
$\langle ...\rangle$ denotes the random matrix average over the
measure (A.1), and ${\bm E}$ and $L$ are $2r N \times 2r N$
matrices
\begin{eqnarray}
{\bm E} &=&
{\rm diag}(
E_{A}^+ {\bm 1}, \overbrace{E_{C}^+ {\bm 1},
\cdots,E_{C}^+ {\bm 1}}^{r-1},
E_{B}^- {\bm 1} , \overbrace{E_{D}^- {\bm 1},
\cdots,E_{D}^- {\bm 1}}^{r-1}) ~, \\
L &=&
{\rm diag}(
\overbrace{{\bm 1}, {\bm 1}, \cdots, {\bm 1}}^{r},
\overbrace{-{\bm 1}, -{\bm 1}, \cdots, -{\bm 1}}^{r}) ~.
\end{eqnarray}
Here ${\bm 1}$ is the $N\times N$ identity matrix. A random matrix
average $\langle ...\rangle$ can be easily evaluated, and we obtain

\begin{eqnarray}
{\cal Z} 
&=& \lim_{r\to 0 }
\int d[\psi] e^{i \psi^{\dagger} L {\bm E} \psi}
\exp \left( -{\lambda^2 \over 4N } {\rm  Tr} a^2
+ {\mu^2 \over 4N} {\rm Tr} b^2 \right) , 
\end{eqnarray}
where $\lambda^2 = (\sigma_{\rm O} - \sigma_{U} )e^{-2 \tau }
+ \sigma_{\rm U}$ and $\mu^2 = \sigma_{U} (1 -e^{-2\tau })$.
The matrices $a$ and $b$ are $N \times N$ matrices, whose elements
are given by
\begin{eqnarray}
a_{\ell , m} &=&
\sum_{\eta}
\psi_{\eta , \ell}^{\ast} \Lambda^{\eta , \eta} \psi_{\eta , m}
+
\sum_{\eta}
\psi_{\eta , m}^{\ast} \Lambda^{\eta , \eta} \psi_{\eta , \ell} ~ , \\
b_{\ell , m} &=&
\sum_{\eta}
\psi_{\eta , \ell}^{\ast} \Lambda^{\eta , \eta} \psi_{\eta , m}
-
\sum_{\eta}
\psi_{\eta , m}^{\ast} \Lambda^{\eta , \eta} \psi_{\eta , \ell} ~ .
\end{eqnarray}
Here $\eta$ is one of the indices $A,C_1,\cdots,C_{r-1},B,
D_1,\cdots , D_{r-1}$. Let us define $\Lambda^{\eta, \eta'}$
as the $(\eta, \eta')$ element of the diagonal matrix
\begin{eqnarray}
\Lambda &=& {\rm diag}(\overbrace{1,1,\cdots,1}^{r},\overbrace{-1,-1,\cdots,-1}^{r}) .
\end{eqnarray}
Next we introduce the vector $\Psi_{\ell}$ and $\bar{\Psi}_{\ell}$
which have $4r $ components as
\begin{eqnarray}
\Psi_{\ell} &=& {1\over \sqrt{2}}
\left(
\begin{array}{c}
\psi_{\ell} \\
\psi^{\ast}_{\ell}
\end{array}
\right) ,
\qquad
\bar{\Psi}_{\ell} = {1\over \sqrt{2}}
\left(
\begin{array}{c}
\psi^{\dagger}_{\ell} ,\psi^{T}_{\ell}
\end{array}
\right)
\end{eqnarray}
Then ${\rm Tr}a^2$ and ${\rm Tr}b^2$ can be written in terms of
these vectors as
\begin{eqnarray}
{\rm Tr} a^2 &=& 4 \sum_{\ell , m }
\bar{\Psi}_{\ell} {\bm M} \Psi_{m} \bar{\Psi}_{m} {\bm M} \Psi_{\ell} , \\
{\rm Tr} b^2 &=& -4 \sum_{\ell , m }
\bar{\Psi}_{\ell} \bar{\bm M} \Psi_{m} \bar{\Psi}_{m}
 \bar{\bm M} \Psi_{\ell} ,
\end{eqnarray}
where ${\bm M}$ and $\bar{\bm M}$ are defined as
\begin{eqnarray}
{\bm M}&=&
\left(
\begin{array}{cc}
\Lambda ,& {\bm 0} \\
{\bm 0}, & \Lambda \\
\end{array}
\right)
, \qquad
\bar{\bm M}=
\left(
\begin{array}{cc}
\Lambda ,& {\bm 0} \\
{\bm 0}, & -\Lambda \\
\end{array}
\right) .
\end{eqnarray}
By introducing $4r \times 4r$ matrices $A$ and $\bar{A}$ as
\begin{eqnarray}
A &=& \sum_{m=1}^{N} \Psi_{m} \bar{\Psi}_{m} {\bm M}, \qquad
\bar{A}
= \sum_{m=1}^{N} \Psi_{m} \bar{\Psi}_{m} \bar{\bm M} ,
\end{eqnarray}
we can write the generating function in the form
\begin{eqnarray}
{\cal Z} 
&=&
\int d[\Psi] e^{i \psi^{\dagger} L {\bm E} \psi}
e^{-{\lambda^2 \over N} ( {\rm Tr} A^2 + \nu^2 {\rm Tr} \bar{A}^2 ) } ,
\end{eqnarray}
where $\nu^2 = (\mu / \lambda)^2$. Here and hereafter we omit the symbol
$\lim_{r\to 0}$.  
In order to write $A$ and $\bar{A}$ in a unified way,
we introduce ${\cal A}$ as
\begin{eqnarray}
{\cal A} &=& b_{+} A + b_{-} \hat{\tau}_3
A \hat{\tau}_3   , \\
b_{\pm} &=& {1\over 2}
\left( \sqrt{1 + \nu^2 } \pm \sqrt{1 - \nu^2 } \right) ,
\end{eqnarray}
where $\hat{\tau}_3$ is
\beq
\hat{\tau}_3 &=&
{\rm diag} (
\overbrace{1, 1, \cdots, 1}^{2 r},
\overbrace{-1, -1, \cdots , -1}^{2 r}) .
\eeq
Notice that $\bar{A}=A \hat{\tau}_3 $, and 
${\rm Tr}{\cal A}^2 = {\rm Tr} A^2 + \nu^2 {\rm Tr}[(A \hat{\tau}_{3})^2 ]$.
Thus, the generating function is written in terms of the
new matrix ${\cal A}$ as
\begin{eqnarray}
{\cal Z} 
&=&
\int d[\psi] e^{i \psi^{\dagger} L {\bm E} \psi}
e^{-{\lambda^2 \over N}  {\rm Tr} {\cal A}^2 }  .
\end{eqnarray}
We then employ the Hubbard-Stratonovich transformation to 
get rid of the  term quadratic in $A$ and
consequently quartic in $\psi$. Using the relation $ {\cal N} \int
d[Q] \exp \left( {N\over 4} {\rm Tr} Q^2 \right) =1, $ where $Q$
is anti-Hermitian, we can readily derive the expression
\begin{eqnarray}
 {\cal Z} 
&=&
{\cal N}\int d[Q] \int d[\psi]
\exp \left\{ i \psi^{\dagger} L {\bm E} \psi
+ {N\over 4} {\rm Tr} Q^2 - \lambda {\rm Tr} ({\cal A} Q)
\right\}. 
\end{eqnarray}
Noting that the term ${\rm Tr} ({\cal A} Q)$ is written as
${\rm Tr} ({\cal A} Q)=
{\rm Tr}\{  A ( b_+ Q + b_- \hat{\tau}_{3} Q \hat{\tau}_{3} ) \}.
$ Hence, defining new matrix $\tilde{Q}$ as
\begin{eqnarray}
\tilde{Q} &=&   b_+ Q + b_- \hat{\tau}_{3} Q \hat{\tau}_{3}  ,
\end{eqnarray}
we obtain
\begin{eqnarray}
 {\cal Z} 
&=&
{\cal N}\int d[Q]
\int d[\psi]
\exp \left\{ i \sum_{m=1}^{N} \bar{\Psi}_{m} {\bm M}
\left( {\bm F} + i  \lambda \tilde{ Q} \right)
\Psi_{m}
+ {N \over 4} {\rm Tr} Q^2
\right\}. \nonumber \\
\label{z_mf_expression}
\end{eqnarray}
Here ${\bm F}$ is a $4r\times 4r$ matrix defined as
\begin{eqnarray}
{\bm F} &=& {\rm diag}(E_A , \overbrace{E_C, \cdots, E_C}^{r-1}, E_B,
\overbrace{E_D, \cdots, E_D}^{r-1}, \nonumber \\
& & E_A,
\overbrace{E_C, \cdots, E_C}^{r-1}, E_B,
\overbrace{E_D, \cdots, E_D}^{r-1} ).
\end{eqnarray}
In order to obtain Eq.(\ref{z_mf_expression}), the relation
$
\psi^{\dagger} L {\bm E} \psi =
{1\over 2} \left(
\psi^{\dagger} L {\bm E} \psi +
\psi^{\dagger} {\bm E} L  \psi
\right) = {1\over 2}
\left(
\psi^{\dagger} L {\bm E} \psi +
\psi^{T} L {\bm E}   \psi^{\ast }
\right)
= \sum_{m=1}^{N} \bar{\Psi}_m {\bm M} {\bm F} \Psi_m
$
was used. Evaluating the $\psi$-integral, we reduce
Eq.(\ref{z_mf_expression}) to
\begin{eqnarray}
{\cal Z} 
&=&
{\cal N}\int d[Q]
\exp\left\{
{N\over 4} {\rm Tr} Q^2  -{N\over 2} {\rm Tr} \ln
( {\bm F} + i\lambda \tilde{ Q} )\right\}. 
\end{eqnarray}
We set the average energy $E$ to the origin
(the band center),
and write ${\bm F}$ as
\begin{eqnarray}
{\bm F} &=& {1 \over 2\pi \bar{\rho}} \tilde{\epsilon} ~ , \\
\tilde{\epsilon } &=&
{\rm diag} (
\epsilon_A,
\overbrace{\epsilon_C, \cdots , \epsilon_C}^{r-1},
\epsilon_B ,
\overbrace{\epsilon_D , \cdots , \epsilon_D}^{r-1},
\epsilon_A,
\overbrace{\epsilon_C, \cdots , \epsilon_C}^{r-1},
\epsilon_B ,
\overbrace{\epsilon_D , \cdots, \epsilon_D}^{r-1} ) ~ , \nonumber \\
\eeq
where the average density is $\bar{\rho} = N/(\pi\lambda )$.
Now we need to consider the effect of $\nu$.
We treat it as a perturbation~\cite{altland93}
and expand $b_+$ and $b_-$ with respect to $\nu$ as
$
b_+ \sim  1 $ and $ b_- \sim  {\nu^2 \over 2}
$.
Setting $\nu =O(N^{-1/2})$, we find a dominant contribution
to the generating function
\begin{eqnarray}
 {\cal Z} 
&=&
{\cal N}\int d[Q]
\exp \left\{
-{N \over 2} {\rm Tr} \ln
 \left( 1 +
{ {\lambda \over 2N}\tilde{\epsilon} +
{i\lambda \nu^2 \over2}\hat{\tau}_{3} Q \hat{\tau}_{3}  \over  i\lambda Q }
\right)
\right\}
\nonumber \\
&& \times
\exp\left[ N
\left\{
{1\over 4} {\rm Tr} Q^2  -{1\over 2} {\rm Tr} \ln
( i\lambda Q )
\right\}
\right] ~ .\label{sd}
\end{eqnarray}
The next step is the saddle point analysis in the limit $N \rightarrow \infty$.
We find the saddle point equation
\begin{eqnarray}
{\delta \over \delta Q_{ij}}
\left\{
{1\over 4} {\rm Tr} Q^2  -{1\over 2} {\rm Tr} \ln
( i\lambda Q )
\right\} &=& 0 ~ .
\end{eqnarray}
This equation yields a saddle point $Q_*$ satisfying
\begin{eqnarray}
Q_*^2 =1 ~. 
\end{eqnarray}
As in the case of the GOE, the saddle point satisfies
the symmetry relation
\beq
\sigma_1 Q_*^{T} \sigma_1 &=& Q_* ~. \label{symmetry1}
\eeq
Here $\sigma_1$ is the Pauli matrix 
$\left( \begin{array}{cc} 0, & {\bf 1} \\ {\bf 1}, & 0 
\end{array} \right)$.
Expanding the logarithmic part of Eq.(\ref{sd}),
we obtain the nonlinear sigma model
\begin{eqnarray}
 {\cal Z} 
&=&
{\cal N}\int d[Q_*]
\exp \left\{
-{N \over 2} {\rm Tr} \ln
 \left( 1 +
{ {\lambda \over 2 N}\tilde{\epsilon} +
{i\lambda \nu^2 \over2}\hat{\tau}_{3} Q_* \hat{\tau}_{3}  \over  i\lambda Q_* }
\right)
\right\} \nonumber \\
&=&
{\cal N}\int d[Q_*]
\exp \left[
 {i \over 4 } {\rm Tr} (\tilde{\epsilon} Q_* )
- {N\nu^2 \over 4} {\rm Tr} \left\{
( Q_* \hat{\tau}_{3} )^2 \right\} \right] \nonumber \\
&=&
{\cal N}\int d[Q_*]
\exp \left[
 {i \over 4 } {\rm Tr} (\tilde{\epsilon} Q_* )
- {s \over 4} {\rm Tr} \left\{ ( Q_* \hat{\tau}_{3} )^2 \right\}  \right] ~.
\label{nls}
\end{eqnarray}
Here we introduced a parameter $s =N\nu^2 $.

\subsection{Parametrization}
We further consider the exponential factor in Eq. (\ref{nls}).
Let us denote the exponent in Eq.(\ref{nls}) by
\beq
{\cal S}(Q_* ) &=&
 {i \over 4 } {\rm Tr} (\tilde{\epsilon} Q_* )
- {s \over 4} {\rm Tr} \left\{ ( Q_* \hat{\tau}_{3} )^2 \right\}  ,
\eeq
As in the GOE case, we parametrize $Q_*$ by using
the matrices $B$ and $B^{\dagger}$ as
\beq
Q_*
&=&
\left\{
{\bm 1} -
\left(
\begin{array}{cc}
0 , & B \\
B^{\dagger} ,& 0
\end{array}
\right)
\right\}
\left(
\begin{array}{cc}
{\bm 1} ,& 0 \\
0 , & - {\bm 1}
\end{array}
\right)
\left\{
{\bm 1} -
\left(
\begin{array}{cc}
0 , & B \\
B^{\dagger} ,& 0
\end{array}
\right)
\right\}^{-1} \nonumber \\
&=&
\left\{
{\bm 1} -
\left(
\begin{array}{cc}
0 , & B \\
B^{\dagger} ,& 0
\end{array}
\right)
\right\}
\left(
\begin{array}{cc}
{\bm 1} ,& 0 \\
0 , & - {\bm 1}
\end{array}
\right)
\sum_{n=0}^{\infty}
\left(
\begin{array}{cc}
0 , & B \\
B^{\dagger} ,& 0
\end{array}
\right)^n  .
\eeq
In this representation, the symmetry relation of
$Q_s$ (\ref{symmetry1}) is satisfied by requiring \beq B^{\dagger}
&=& - \sigma_1 B^{T} \sigma_1 \label{symmetry2}. 
\eeq

From a straightforward calculation, we obtain the expression of
${\cal S}(B ) ~(\, = {\cal S}(Q_* )\,)$ in the form 
\beq {\cal
S}(B )&=&
{i\over 2} (\epsilon_A - \epsilon_B - \epsilon_C + \epsilon_D ) \nonumber \\
&+& {i\over 2} {\rm Tr} (\tilde{\epsilon}_{AC} B B^{\dagger } )
- {i \over 2} {\rm Tr} ( \tilde{\epsilon}_{BD}  B^{\dagger } B )
-2s {\rm Tr}(B B^{\dagger}) +2s {\rm Tr} (B\tau_3 B^{\dagger} \tau_3 )
\nonumber \\
&+& {i \over 2} \sum_{m=2}^{\infty}
\left[
{\rm Tr} \left\{ \tilde{\epsilon}_{AC} (B B^{\dagger } )^m \right\}
-
{\rm Tr} \left\{ \tilde{\epsilon}_{BD} ( B^{\dagger } B )^m \right\} \right]
-2s \sum_{m=2}^{\infty} {\rm Tr} (B B^{\dagger})^m
\nonumber \\
&-& s \sum_{m=1}^{\infty } \sum_{m'=1}^{\infty } {\rm Tr}
\left\{ (B B^{\dagger} )^m \tau_3 ( B B^{\dagger})^{m'} \tau_3 \right\}
\nonumber \\
&-& s
\sum_{m=1}^{\infty } \sum_{m'=1}^{\infty } {\rm Tr}
\left\{ (B^{\dagger} B)^m \tau_3 ( B^{\dagger} B)^{m'} \tau_3 \right\}
\nonumber \\
&+&2s \hspace*{-1cm}
\sum_{m,m'=0 ; (m,m')\ne (0,0)}^{\infty} \hspace*{-1cm}
{\rm Tr} \left\{ (B B^{\dagger})^m B \tau_3 (B^{\dagger} B)^{m'} B^{\dagger}
\tau_3 \right\} \, ,
\eeq
where $\tilde{\epsilon}_{AC,BD}$ and $\tau_3$ are $2r \times 2r$ matrices
defined in (\ref{epsac}), (\ref{epsbd}) and (\ref{tau3}).


\section*{References}
\begin{thebibliography}{10}
\bibitem{haake}
F. Haake, {\em Quantum Signatures  of Chaos}, 2nd ed.
(Springer, Berlin, 2001).
\bibitem{bgs}
O. Bohigas, M. J. Giannoni, and C. Schmit, Phys. Rev. Lett. {\bf 52}, 1 (1984);
S. W. McDonald and A. N. Kaufmann, Phys. Rev. Lett. {\bf 42}, 1189 (1979);
G. Casati, F. Valz-Gris, and I. Guarneri, Lett. Nuovo Cim. {\bf 28}, 279 (1980);
M. V. Berry, Proc. R. Soc. Lond. A {\bf 413}, 183 (1987).
\bibitem{muller}
S. M\"{u}ller, S. Heusler, P. Braun, F. Haake, and A. Altland, Phys. Rev. Lett.
{\bf 93}, 014103 (2004), Phys. Rev. E {\bf 72}, 046207 (2005).
\bibitem{berry}
M. V. Berry, Proc. R. Soc. Lond. A {\bf 400}, 229 (1985).
\bibitem{sr}
M. Sieber and K. Richter, Physica Scripta {\bf T90}, 128 (2001);
M. Sieber, J. Phys. A: Math. Gen. {\bf 35}, L613 (2002).
\bibitem{PM83}
A. Pandey and M.L. Mehta, Commun. Math. Phys. {\bf 87} 449 (1983).
\bibitem{bgoas}
O. Bohigas, M.-J. Giannoni, A.M. Ozorio de Almeida and C. Schmit,
Nonlinearity {\bf 8}, 203 (1995).
\bibitem{nagaosaito2003}
T. Nagao and K. Saito, Phys. Lett. A {\bf 311}, 353 (2003).
\bibitem{tr}
M. Turek and K. Richter, J. Phys. A: Math. Gen. {\bf 36}, L455 (2003).
\bibitem{saitonagao2006}
K. Saito and T. Nagao, Phys. Lett. A {\bf 352}, 380 (2006).
\bibitem{nagaohaake2007}
T. Nagao, P. Braun, S. M\"{u}ller, K. Saito, S. Heusler, and F. Haake,
J. Phys. A: Math. Theor. {\bf 40}, 47 (2007).
\bibitem{sieber2007}
J. Kuipers and M. Sieber, J. Phys. A: Math. Theor. {\bf 40}, 935 (2007).
\bibitem{nagaosaito2007}
T. Nagao and K. Saito, J. Phys. A: Math. Theor. {\bf 40},
12055 (2007).
\bibitem{heusler2007}
S. Heusler, S. M\"{u}ller, A. Altland, P. Braun, and F. Haake,
Phys. Rev. Lett. {\bf 98}, 044103 (2007),
S. M\"{u}ller, S.Heusler, A.Altland, P.Braun, and F.Haake, arXiv:0906.1960.
\bibitem{Zirnbauer}
Zirnbauer M 2009 private communication
\bibitem{berry_keating90}
M. V. Berry and J. P. Keating, J. Phys. A: Math. Gen. {\bf 23}, 4839 (1990),
J. P. Keating, Proc. R. Soc. Lond. A {\bf 436}, 99 (1992), M. V. Berry
and J. P. Keating, Proc. R. Soc. Lond. A {\bf 437}, 151 (1992).
\bibitem{keating_muller}
J. P. Keating and S. M\"{u}ller, Proc. R. Soc. Lond. A {\bf 463} 3241 (2007).
\bibitem{HOdA}
J.H. Hannay and A.M. Ozorio de Almeida, J. Phys. A: Math. Gen. {\bf 17}, 3429 (1984).
\bibitem{efetov}
K. Efetov, Supersymmetry in Disorder and Chaos, (Cambridge University Press, 1997).
\bibitem{altland93}
A. Altland, S. Iida, and K. B. Efetov, J. Phys. A: Math. Gen. {\bf 26}, 3545 (1993).
\end{thebibliography}
\end{document}